\documentclass[11pt]{JHEP3}
\usepackage{amsmath,epsfig}
\usepackage{amssymb,amsfonts}
\usepackage{datetime}
\usepackage{mathrsfs}
\usepackage{latexsym}
\usepackage{multirow,cite}
\newcommand{\non}{\nonumber}
\def\p{\partial}
\def\a{\alpha}
\def\b{\beta}
\def\d{\delta}
\def\g{\gamma}
\def\l{\lambda}
\def\e{\epsilon}
\def\k{\kappa}
\def\th{\theta}
\def\om{\omega}
\def\r{\rightarrow}
\def\Om{\Omega}
\def\S{\Sigma}
\def\G{\Gamma}
\def\half{\frac{1}{2}}
\def\A{\mathcal{A}}
\def\B{\mathcal{B}}
\def\O{\mathcal{O}}
\def\s{\sigma}
\newcommand{\bea}{\begin{eqnarray}}
\newcommand{\eea}{\end{eqnarray}}
\newcommand{\be}{\begin{equation}}
\newcommand{\ee}{\end{equation}}
\newcommand{\bi}{\begin{itemize}}
\newcommand{\ei}{\end{itemize}}

\title{Stringy Schr\"{o}dinger truncations}
\date{}
\author{St\'ephane Detournay$^\dag$ and Monica Guica$^\ddag$\\

\vspace{1mm}

\hspace{-5mm}{\it $^\dag$ Center for the Fundamental Laws of Nature, Harvard University, \\
 \hspace{-0.25 cm} Cambridge, MA 02138, USA} \\
\vspace{-1mm}

\hspace{-5mm}{\small  ${}^\ddag$ David Rittenhouse Laboratory, University of Pennsylvania, \\  \hspace{-0.33 cm}  Philadelphia, PA 19104, USA}}

\abstract{ \bigskip

Motivated by the desire to better understand finite-temperature holography for three-dimensional Schr\"{o}dinger spacetimes, we: i) construct a four-parameter family of warped black string solutions of type IIB supergravity and  ii) find the first consistent truncations of type IIB string theory to three dimensions that admit both supersymmetric Schr\"{o}dinger solutions and warped generalizations of the BTZ black hole.

\medskip

Our analysis reveals a number of interesting features. One is that the thermodynamic properties of all the warped black strings, as well as the asymptotic symmetry group data, are identical to those of BTZ, in an appropriate parametrization. A  more striking feature is that the spectrum of linearized perturbations around the various supersymmetric Schr\"{o}dinger vacua  oftentimes contains modes that carry energy flux through the spacetime boundary, which are usually  believed to be unstable.
A preliminary analysis indicates that, at least in the case of most interest, these modes do not lead to an instability.

\medskip

}

\begin{document}

\section{Introduction}

Two-dimensional conformal field theories appear to play a special role in the microscopic understanding of black hole entropy. For example, all black holes that exhibit an $AdS_3$ factor in their near-horizon region, including the one originally studied by Strominger and Vafa \cite{Strominger:1996sh}, are microscopically described by two-dimensional CFTs\cite{Strominger:1997eq}. This universal fact is reflected in the equality of the Bekenstein-Hawking and Cardy entropies, in the agreement of black hole greybody factors and CFT thermal Green's functions \cite{Maldacena:1996ix, Maldacena:1997ih, Cvetic:1997uw}, in the spectrum of quasinormal modes \cite{Birmingham:2001pj}, and so on.

However, black holes with a near-horizon $AdS_3$ factor 
are only  a very specific subclass of all known solutions. Recently, the Kerr/CFT proposal \cite{kerrcft}  and its generalizations  asserted that general \emph{extremal} black holes, including certain astrophysical examples, are also described by two-dimensional CFTs. This proposal  seemed surprising at first, given that the near-horizon geometry 
 of generic extremal black holes only has  $SL(2,\mathbb{R}) \times U(1)$ isometry, and the universal factor \cite{
Bardeen:1999px, Kunduri:2007vf,Kunduri:2008rs} that all geometries contain is not $AdS_3$, but a  deformed version of it known as warped $AdS_3$. Despite this, the evidence in favour of this conjecture seemed plentiful: the presence of a Virasoro symmetry that enhances the $U(1)$ isometry factor, the precise
agreement of the Bekenstein-Hawking entropy of the black hole with the Cardy entropy
of the putative dual CFT, the striking resemblance between scattering amplitudes off the
black hole and thermal CFT Green's functions \cite{scattampl, Cvetic:2009jn, Hartman:2009nz, Chen:2010bsa}.

Several arguments also linked  general non-extremal black holes to conformal field theories, based on the approximate  conformal symmetry of the wave equation for test fields in a general black hole background \cite{Castro:2010fd}.
These symmetries have been realized geometrically through the introduction of subtracted geometries \cite{Cvetic:2011hp, Cvetic:2011dn, Cvetic:2012tr}, which have lead to a very interesting recent proposal that non-extremal black holes are described by certain irrelevant deformations of a CFT \cite{Baggio:2012db}. In this article nevertheless, we will only concentrate on extremal and near-extremal black holes, since to date they are the better understood.




In spite of the rather convincing evidence we quoted, the relationship between extremal black holes and two-dimensional CFTs turns out to be rather subtle and complicated. The 
Kerr/CFT holographic dictionary indicates that the conformal dimensions of operators  depend on the momentum along one particular direction. This feature is at odds with the assumption that the dual microscopic theory is a CFT - since then the operator dimensions would have to be constants - as well as with locality.  Moreover, the momentum-dependent conformal dimensions can at times become imaginary \cite{Dias:2009ex,nodyn}, a feature that seems incompatible with the unitarity of the proposed dual CFT description.

 In order to clarify these issues, it has been useful to study one particular instance of the Kerr/CFT correspondence using string theory \cite{microkerr}. 
In this simpler example, it was argued \cite{kerrdip} that the question of understanding  Kerr/CFT  is   equivalent to understanding finite-temperature holography in a particular three-dimensional Schr\"{o}dinger spacetime. As a reminder, the metric of the vacuum $Sch_3$ spacetime is

\be 
ds^2 =- \l^2 r^2 du^2 + 2 r du dv + \frac{dr^2}{4r^2} 
\ee 
Schr\"{o}dinger spacetimes are notoriously \emph{not} dual to relativistic conformal field theories. Rather, their holographic dual can be understood as a particular irrelevant deformation of a relativistic CFT \cite{nr}. This deformation  preserves non-relativistic conformal invariance and introduces non-localities at the scale $\l$ in the $v$ direction\footnote{Both features are present in the only known explicit example of a field theory dual to a Schr\"{o}dinger spacetime, the so-called dipole theory \cite{
Bergman:2000cw,Bergman:2001rw}.}, which are reflected in the characteristic momentum-dependent conformal dimensions of operators. Upon compactifying the null direction $v$, the dual theory is conjectured to reduce to a strongly-coupled one-dimensional non-relativistic CFT \cite{Son:2008ye, Balasubramanian:2008dm}.

The connection between Kerr/CFT and Schr\"{o}dinger holography shows that the dual theory is not a local theory; in particular, it is not a relativistic CFT. Nevertheless, it does not explain the emergence of an apparently local gravity description from this non-local theory, the applicability of Cardy's formula for counting the black hole entropy, and the appearance of a \emph{relativistic} spacetime  conformal symmetry group from a boundary theory which only exhibits $SL(2,\mathbb{R}) \times U(1)$ \emph{non-relativistic} conformal invariance. 

An interesting proposal for how  to circumvent the last  puzzle has been put forth in  \cite{Detournay:2012pc}. The authors studied the hypothetical \emph{warped conformal field theories} first described in \cite{Hofman:2011zj}, which are two-dimensional local quantum field theories with $SL(2,\mathbb{R}) \times U(1)$ symmetry  enhanced to a direct product of a Virasoro and a Ka\v{c}-Moody algebra. These theories  exhibit an asymptotic growth of states  very similar to the Cardy growth of usual two-dimensional CFTs,  which has been shown to correctly reproduce the Bekenstein-Hawking entropy of various black holes in warped $AdS_3$.
This approach has the potential to explain the entropy of extreme back holes in terms of non-local deformations of warped CFTs
without invoking the need for relativistic conformal symmetry; if this is the case, then the boundary conditions that  define near-horizon dynamics and symmetries of the extreme Kerr black hole have to be completely reconsidered, along lines similar to those of \cite{Castro:2009jf}.


In view of this new option, the question that we would like to answer is: what are the natural boundary conditions to impose on metric perturbations in Schr\"{o}dinger - and by extension NHEK - spacetimes?  Are they the ones that yield a \emph{relativistic} CFT structure - similar to those originally proposed in \cite{kerrcft} - or the ones where the emerging structure is that of a \emph{ non-relativistic} ``warped CFT''? Are both sets of boundary conditions sensible? If yes, what is the difference - if any - between the boundary field theories they define?

It should be possible to give a precise answer to the above questions within the rather constraining framework of holographic renormalization. While several  careful studies of holographic renormalization for three-dimensional Schr\"{o}dinger spacetimes, with an emphasis on the stress tensor sector, already exist, consensus has not yet been reached: on the one hand, the perturbative analysis of \cite{vanRees:2012cw} finds a non-relativistic stress-tensor complex that  naturally generates non-relativistic CFT symmetries (only the $SL(2,\mathbb{R})$ isometry is enhanced to a Virasoro), whereas the Fefferman-Graham-like expansion of \cite{fg} suggests a strong connection to relativistic CFT$_2$s and the possibility to construct two sets of Virasoro asymptotic symmetry generators. Other analyses performed in different spacetime dimensions from different points of view include \cite{Ross:2009ar,Hartong:2010ec}. Nevertheless, to date it is unclear which are the correct holographic variables to use, with their associated boundary conditions.

In order to develop the full holographic dictionary in three-dimensional Schr\"{o}dinger spacetimes, one needs a good toy model. In particular, if one is interested in finite-temperature holography and the stress tensor sector, this toy model  
should contain black holes. One of the main drawbacks of the theories previously analysed in the literature - such as 
 Einstein gravity  coupled to  a massive vector - is the fact that they do not contain any (known) black holes.\footnote{Three-dimensional Schr\"{o}dinger holography has also been studied for the case of topologically massive gravity \cite{Anninos:2010pm,nr}, a theory which does contain black hole solutions \cite{Nutku:1993eb,Moussa:2003fc,Anninos:2008fx}. Nevertheless, the theory appears to be non-unitary \cite{Skenderis:2009nt,Andrade:2009ae} and it is not clear whether holography should make sense. 
}. Thus, one cannot perform important tests of the proposed holographic stress tensor or address thermodynamic questions - such the origin of 
 the Cardy formula - from a holographic point of view.

The main aim of this paper  is to fill in this gap by  providing a set of simple three-dimensional models of gravity coupled to massive vector fields and scalars that admit supersymmetric Schr\"{o}dinger solutions, as well as warped generalizations of the BTZ black hole. 
All these models are  obtained as consistent truncations of type IIB supergravity to three dimensions. They are simple enough that the linearized equations of motion above the Schr\"{o}dinger backgrounds can be solved exactly using Mathematica, allowing us to probe certain peculiar features of stringy Schr\"{o}dinger spacetimes, which we will discuss shortly.

A secondary aim is to understand the geometric structure of finite-temperature states above Schr\"{o}dinger spacetimes more generally. To this end, we construct  a four-parameter family of warped six-dimensional black string solutions of type IIB supergravity, supported by Ramond-Ramond flux. Two of the four parameters represent the left/right temperatures associated with the black strings, whereas the other two label the various type IIB Schr\"{o}dinger-invariant vacua. Interestingly, we find that - with an appropriate parametrization - the mass, linear momentum, entropy and asymptotic symmetry group data 
 are identical to those of the BTZ black string with the same left/right-moving temperature. In particular, they are all independent of the two warping parameters.

Despite the fact that the general solution for the black strings is rather involved, the three-dimensional non-compact part of these geometries takes a surprisingly similar form for all warping parameters, if  the temperatures are kept fixed. Thus, we may refer to a universal, ``warped'' analogue of the BTZ black hole, which is  the same across all solutions. These three-dimensional black holes seem to fit inside the three-dimensional consistent truncations only in very special cases, which fortunately are the ones that interest us the most. It seems nevertheless surprising that in order to study finite-temperature holography for the general backgrounds, one is forced to work in six dimensions.

Finally,  the consistent truncations can be used to find (a subset of) the linearized perturbative spectrum of $SL(2,\mathbb{R})$ conformal dimensions of type IIB supergravity about the various Schr\"{o}dinger backgrounds. Despite the fact that in all the examples we study, the vector field $\p_u$ is always timelike, many of the spectra turn out to have a rather surprising feature: they contain the so-called ``travelling wave'' solutions, which are modes that  have oscillatory behaviour at infinity and carry flux through the spacetime boundary. The AdS analogues of these modes are fields below the Breitenlohner-Friedman bound, which are known to trigger an instability of the spacetime. Nevertheless, whether this instability exists in Schr\"{o}dinger spacetimes is not clear  \cite{Hartnoll:2008rs,moroz,Blau:2010fh}. A preliminary analysis that we performed at linearized level, did not signal its presence.  Nevertheless, one should also check if there exists a non-linear instability, as predicted in \cite{Hartnoll:2008rs}. It would be very interesting if these modes did have a holographic interpretation, as they may shed light on the phenomenon of near-horizon superradiance.



\bigskip

This paper is structured as follows. In section \ref{nwbssf}, we explain how to generate the four-parameter family of black string solutions of type IIB supergravity, using 
generalized spectral flows. In particular, in \ref{spci} 
we fix the terminology that we will be using throughout the paper. In section \ref{physprop}, we show that the newly-found black string solutions satisfy the first law of thermodynamics and we perform various asymptotic symmetry group analyses of these spacetimes. 
In section \ref{seccttr}, we present several consistent truncations, organised by the various cases of interest. In section \ref{strspec}, we analyse the qualitative features of the spectra, singling out the cases that admit travelling wave solutions.  In section \ref{stabanal} we perform the linearized stability analysis in presence of the travelling waves for one supersymmetric background of particular interest, and find no instability. Many of the  details of our analyses can be found in the appendices.

\section{New warped black strings from spectral flow \label{nwbssf}}

Schr\"{o}dinger spacetimes in string theory are oftentimes constructed by starting with an anti-de Sitter solution and acting on it with various solution-generating transformations, followed by an infinite boost 
\cite{Alishahiha:2003ru}. Black holes in Schr\"{o}dinger spacetimes are similarly obtained by starting with an AdS black hole and applying the same solution-generating transformation, followed by a boost.

The  Schr\"{o}dinger spacetimes best studied so far are the ones dual to the so-called dipole theories \cite{Bergman:2000cw,Bergman:2001rw}. These backgrounds are generated by a solution-generating transformation known as TsT (described below). Just like $AdS_{d+1} \times S^p$, the solution is supported by self-dual Ramond-Ramond flux; additionally, it carries nonzero NS-NS  B-field with one leg along the Schr\"{o}dinger factor and one leg along the sphere. The metric is a direct product of the two factors.

The main focus of this paper will be Schr\"{o}dinger backgrounds - and the corresponding black holes - in type IIB string theory that are supported entirely by Ramond-Ramond flux. (Nevertheless, we will often include the usual dipole backgrounds in our discussion, for completeness). The full  spacetimes  will in general not be a direct product of a sphere and a Schr\"{o}dinger factor. We construct these backgrounds  by 
 applying the solution generating techniques of \cite{Bena:2008wt,stubena} to the near-extremal D1-D5 black string solution of type IIB string theory. They are parametrized  by two thermodynamic potentials - the left/right moving temperatures $T_\pm$ - of the orginal BTZ black string  and two deformation parameters: the parameters of the two generalized spectral flows\footnote{
 These solutions differ from those studied in \cite{stubena} by the fact that we allow  two arbitrary temperatures, as opposed to a single one, and in the more intuitive parametrization.}.

We start this section by reviewing how  the three-dimensional dipole black hole solution in type IIB  string theory is generated. Next, in \ref{nwbh}, we explain how to generate similar but new warped black holes in string theory, using the solution generating techniques of \cite{stubena}. In \ref{spci}, we write down the explicit solutions in several cases of interest, for quick reference and in order to fix the terminology.

\subsection{Review of dipole black strings \label{dipbh} }

The prototype of a string-theoretical black hole in warped $AdS_3$ is the finite temperature dipole background. This is obtained by starting with the near-horizon geometry of the D1-D5 black string solution

\bea
ds^2 & = & - \frac{(\rho^2 - \rho_+^2)(\rho^2 - \rho_-^2)}{\rho^2} d\tau^2 + \frac{\ell^2\rho^2 d\rho^2}{(\rho^2 - \rho_+^2)(
\rho^2 - \rho_-^2)}
+ \rho^2 \left(R \, d x^5 - \frac{\rho_+ \rho_-}{\rho^2}d\tau\right)^2 +\non \\
&& \hspace{1 cm} + \frac{\ell^2}{4} (d\theta^2 + \sin^2 \th d\phi^2) + \frac{\ell^2}{4} (d\psi+\cos \th d\phi)^2 +  \sum_i dx_i^2 \label{bssol}
\eea
with RR two-form potential and dilaton
\be
C^{(2)} =   \frac{Q}{4}  \cos \th d\phi \wedge d\psi -  R \rho^2 d\tau \wedge d\,x^5 \;, \;\;\;\;\; e^{-2 \Phi} = 1 \label{btztwof}
\ee
For simplicity, we have set $Q_1=Q_5 = Q = \ell^2$. The coordinate $x^5$ is identified mod $2\pi$.  This black string background corresponds to a thermal ensemble in the dual D1-D5 CFT, characterized by the dimensionless left/right-moving temperatures $T_{L,R}$, where

\be
 \rho_\pm = \pi \ell  (T_R \pm T_L)
\ee
The dipole background is obtained by applying a solution generating transformation known as TsT\footnote{Note that in the decoupled geometry we have aready sent $\a'$ to zero. Nevertheless, TsT as a block acts solely within the context of IIB supergravity, and we will be using it as such.}, which consists of a T-duality on $x^5$, a shift $\psi \r \psi + 2 \l \tilde x^5$ , and then a T-duality back on $\tilde x^5$, where $\tilde x^5$ is the coordinate T-dual to $x^5$. What we are interested in is the null dipole background - which contains a Schr\"{o}dinger factor - and can be obtained  by performing an infinite boost on the $\tau,x^5$ coordinates, accompanied by an appropriate focusing of the energies/ rescaling of the temperatures. All the steps have been worked out in detail in section $6$ of \cite{kerrdip}, and also appear in \cite{stromwei}.

The solution that one obtains after the TsT is somewhat complicated, but the infinite boost simplifies it drastically. Since the two procedures commute - at least at the level of the supergravity solution - a much simpler way to arrive at the same final supergravity background is to first perform the infinite boost, and only then the TsT transformation. Let us define

\be
u = R \, x^5+ \tau \;, \;\;\;\;\;  v = R \, x^5 - \tau \;, \;\;\;\;\; r = \frac{1}{2\ell^2}\left[ \rho^2 -\half ( \rho_+^2+ \rho_-^2) \right] 
\ee

\be
 T_+ = \pi  T_R\;, \;\;\;\;\; T_-=   \pi T_L 
\ee
In these coordinates, the metric and two-form potential take the following simple form

\be
ds^2 = \ell^2 \left(T_-^2 du^2 + T_+^2 dv^2 + 2 r du dv  + \frac{ dr^2}{4 (r^2-  T_+^2 T_-^2)} +  d\Om_3^2 \right)\label{bsnull}
\ee

\be
C^{(2)} =   \frac{\ell^2}{4}  \cos \th d\phi \wedge d\psi -\ell^2 \, r\, du \wedge dv \label{cfnull}
\ee
The $C^{(2)}$ field differs from \eqref{btztwof} by a gauge transformation. The  identifications of the new coordinates are
\be
u \sim u + 2 \pi R \;, \;\;\;\;\; v \sim v +  2\pi R
\ee
The boost and the rescaling of energies that yield a lightlike dipole background are

\be
 u \r e^{\g} u \;, \;\;\;\;\;  v \r e^{-\g} v\;, \;\;\;\; T_\pm \r e^{\pm \gamma} T_\pm\;, \;\;\;\;\; \g \r \infty
\ee
with $r$ kept fixed. Under these rescalings, the form of the metric and $C^{(2)}$-field are invariant, only the coordinate identifications change to

\be
u \sim u + 2 \pi R e^{-\g} \;, \;\;\;\;\; v \sim v + 2 \pi R \,e^{\g} 
\ee
If we take $\g \r \infty$ with $R$ fixed, then the $u,v$ coordiates decompactify. Nevertheless, in order to perform the TsT transformation, which involves a T-duality along the $v$ direction, we need this coordinate to be compact. This can be achieved by taking $\g \r \infty$ with $\tilde R \equiv R \, e^\g $ kept fixed.  Since the original $R \r 0$, this procedure implements the DLCQ of the dual  theory. 
The gravity solution dual to the vacuum $(T_\pm=0)$ exhibits a compact null circle of radius $\tilde R$, and thus the gravity approximation breaks down. This is not what we want, so at the end of our manipulations, we will send $\tilde R \r \infty$,  obtaining a spacetime dual to an unconstrained theory. 

Next, we perform the TsT transformation, which now consists of a T-duality along $v$, a shift $\psi \r \psi + 2 \l \ell^{-2} v'$, and a T-duality back on $v'$. The end result is

\bea
ds^2_{str} &=& \ell^2 \left[\frac{T_+^2 dv^2}{1+ \l^2 T_+^2} + \frac{2 r \, du\, dv}{1+ \l^2 T_+^2}  + du^2 \left( T_-^2 - \frac{\l^2 r^2}{1+ \l^2 T_+^2}  \right) + \frac{ dr^2}{4 \,(r^2 -  T_+^2 T_-^2)}  + \right.\non \\
&& \hspace{2 cm}\left.+ \frac{\s_3^2}{4 (1+ \l^2 T_+^2)} +
\frac{1}{4}\, d\Omega_2^2 \right] + \sum_i dx_i^2 \label{dipbs}
\eea

\be
B = \frac{\l \ell^2}{2 (1 + \l^2 T_+^2)} \left( T_+^2 dv + r du \right) \wedge \s_3 \;, \;\;\;\;\;\; e^{-2\phi} = 1 + \l^2 T_+^2
\ee
which is identical to the one in \cite{kerrdip}, but has taken us considerably less work to obtain. We are using the following shorthands:
\be
d\Om_2^2 = d\th^2 + \sin^2 \th d\phi^2 \;, \;\;\;\;\; \s_3 = d \psi + \cos \th d\phi \;, \;\;\;\;\; d\Om_3^2 = \frac{1}{4} (d\Om_2^2 + \s_3^2)
\ee
Here $d\Om_k^2$ stands for the metric on the unit $S^k$. The $C^{(2)}$-field is still given by \eqref{cfnull}. Salient features of this background are that it is supported by the non-trivial cross B-field. If we set $T_\pm =0$, we obtain the dipole three-dimensional Schr\"{o}dinger background, with

\be
ds^2 = \ell^2\left(2 r du dv - \l^2 r^2 du^2 + \frac{dr^2}{4r^2}\right) + \ell^2 d\Omega_3^2  + \sum_i dx_i^2\;, \;\;\;\;\;
B = \frac{\ell^2}{2} \,\l r du \wedge \s_3 \label{dipsch}
\ee

\be  
C^{(2)} = \frac{\ell^2}{4} \cos \th d\phi \wedge d \psi - \ell^2 \,r du \wedge dv\;, \;\;\;\;\;\; \phi=0
\ee

\subsection{Generating the new black strings \label{nwbh}}

We are interested in  more general Schr\"{o}dinger backgrounds, where there do exist cross terms in the metric, and the RR field is modified from its background value. In \cite{stubena} it has been shown that such warped backgrounds can be generated by applying the so-called generalized spectral flow transformations $\mathscr{T}_{1,2}$ on a seed $AdS_3\times S^3 \times T^4$ solution. To perform these transformations, one writes $AdS_3$ as a $U(1)$ fibre over $AdS_2$ - call it $v$ - and $S^3$ as a $U(1)$ fibre over $S^2$ - call it $\psi$. The spectral flows then mix the Kaluza-Klein vector fields associated to the two fibres in various ways, generally producing a warped $AdS_3$ background. The mixing due to the generalized spectral flows  $\mathscr{T}_1$ and $\mathscr{T}_2$ can be  represented via the following sequence of pseudo-dualities\footnote{These transformations are pseudo-dualities, rather than string theory dualities, because the shift parameters $\hat \l_i$ are not required to be quantized. Nevertheless, they are symmetries of the supergravity action, taking solutions into solutions.}

\bigskip
\begin{center}

\begin{tabular}{rl}
$\mathscr{T}_1$ ( $ STsTS $ ) : & \hspace{1.2 mm}  a type IIB S-duality \\ &
- a T-duality along $v$  \\ &
 - a shift: $\psi \r \psi + 2 \hat \l_1 \tilde v$ \\ &
- a T-duality back on $\tilde v$ \\ &
- a type II B S-duality back
\end{tabular}

\bigskip

\begin{tabular}{rl}
$\mathscr{T}_2$ ( $ T^4STsTS T^4 $ ) : & \hspace{1.2 mm}  four T-dualities along the $T^4$ directions \\ &
- a $\mathscr{T}_1$ transformation, with parameter $\hat \l_2$ \\ &
- four T-dualities on $T^4$ back
\end{tabular}

\end{center}
\bigskip

\noindent The spectral flows can  easily be implemented also if the internal space is $K3$ rather than $T^4$; then the four T-dualities used in the second spectral flow is replaced by the string duality that realizes six-dimensional electromagnetic duality in type IIB compactified on $K3$. The action of generalized spectral flows on a given type IIB geometry has been described in detail in section $3$ of \cite{stubena}.

 If one applies these transformations to a pure RR background such as the D1-D5 string solution, one is sure to only generate RR field. Moreover, cross terms in the metric will generally be induced. 
Applying either generalized spectral flow to $AdS_3 \times S^3 \times T^4$ followed by an infinite boost will generically yield a three-dimensional Schr\"{o}dinger spacetime times the compact factors.  If we subsequently apply the two transformations, $\mathscr{T}_1(\l_1)$ and $\mathscr{T}_2(\l_2)$, we naturally obtain a two-parameter family of Schr\"{o}dinger backgrounds, labeled by the spectral flow parameters $\l_{1,2}$

\be
ds^2 = \ell^2 \left(2 r du dv - \tilde{\l}^2 r^2 du^2 + \frac{dr^2}{4r^2} + \frac{1}{4} \,d\Omega_2^2 + \frac{1}{4}\, \left(\s_3 + \frac{2}{\ell} A\right)^2 \right) + \sum_i dx_i^2 \label{genschm}
\ee
where

\be
\tilde\l = 2 \sqrt{\l_1^2-\l_1 \l_2 + \l_2^2} \;, \;\;\;\;\; A=- 2 \ell (\l_1-\l_2) \, r du \label{tl}
\ee
The backgrounds are supported by purely RR flux 

\be
C^{(2)}  =  \frac{ \ell^2}{4}  \cos \th d\phi \wedge d\psi -   \ell^2 \, r du \wedge d v +    \l_1  \ell^2\,   r du  \wedge \s_3
\ee
and the dilaton vanishes. 

If instead we start with a seed BTZ solution and apply the spectral flows followed by an infinite boost, we obtain a four parameter family of ``warped BTZ'' black strings,  labeled by the spectral flow parameters $\l_1, \l_2$ and the original black hole potentials $T_\pm$. As in the dipole case we have prevously discussed, these solutions are more easily obtained by starting with the boosted metric \eqref{bsnull} and $C^{(2)}$-field \eqref{cfnull}. 
to which we apply the spectral flow transformations reviewed in appendix \ref{genform}.  After a series of coordinate and parameter redefinitions described in detail in appendix \ref{horror}, we find the following solutions\footnote{Note that the parameters $\ell$ and $\l_i$ that appear in  the formulae above are different from the original AdS radius and the shift parameters of the spectral flows. Also, the identification of the squashed sphere coordinate $\psi \sim \psi + 4 \pi$ is not the same as the original identification of the Hopf fibre $\psi$; for details see appendix \ref{horror}.
Nonetheless, the temperatures $T_\pm$ are the same as in \eqref{bsnull}.   }

\be
ds^2  =  \frac{ \ell^2 \sqrt{\S_1 \S_2}}{ (1-\l_2^2 T_+^2)} \left[   \frac{1 }{1+\tilde \l^2 T_+^2} ds_{3}^2  + \frac{1}{4}\, d \Omega_2^2 + \frac{\Delta}{4\S_1 \S_2} \left(\s_3 +\frac{2}{\ell} A\right)^2 \right] + \sqrt{\frac{\S_1}{\S_2}} \, \sum_i dx_i^2 \label{genbhm}
\ee
The above metric is in ten-dimensional string frame and

\be
ds_3^2 =  T_+^2 d v^2 + 2  r du d v  + (1+ \tilde \l^2 T_+^2) T_-^2 du^2 - \tilde \l^2  r^2 du^2 +  (1+ \tilde \l^2 T_+^2) \frac{d r^2}{4( r^2 - T_+^2 T_-^2)}  \label{ref3dm}
\ee

\be
\S_{i} = 1+ \l_i^2  T_+^2 \;, \;\;\;\;\; \Delta = \left(1- \l_1 \l_2 T_+^2  \right)^2 + T_+^2 (\l_1-\l_2)^2 
\ee
The Kaluza-Klein vector field reads

\be
A= - \frac{2 \ell (\l_1 - \l_2)  (1-\l_1 \l_2 T_+^2)}{\Delta} \left( r du + T_+^2 d v \right) \label{akk}
\ee
while the RR two-form and the dilaton are

\be
C^{(2)}  =  \frac{ \ell^2}{4}  \cos \th d\phi \wedge d\psi -  \ell^2\, r du \wedge d v +  \frac{\ell^2  \l_1  }{ \S_1}  \left( r du + T_+^2 d v \right) \wedge \s_3 \;, \;\;\;\;\; e^{2\Phi} = \frac{\S_1}{\S_2} \label{genbhcf}
\ee
Finally, the parameter $\tilde \l$ that enters the three-dimensional metric is given by

\be
\tilde \l =\frac{ 2 \sqrt{\Lambda}}{(1-\l_1^2 T_+^2)(1-\l_2^2 T_+^2)}\;,  \non
\ee

\be
\Lambda = (\l_1^2-\l_1 \l_2 + \l_2^2)(1+ \l_1^2 \l_2^2 T_+^4) - \l_1 \l_2 T_+^2 (\l_1^2 + \l_2^2) \label{tlbh}
\ee
Given the rather complicated dependence of the above solutions on $\l_{1,2}$, it is somewhat surprising that the three-dimensional parts of the above metrics  are so similar to each other, as a function of $T_\pm$. This similarity hints towards the existence of an ``universal'' warped analogue of the BTZ black hole in string theory, which is roughly constant across the various backgrounds/theories we study. Note nevertheless that the remaining fields supporting these backgrounds, such as the Kaluza-Klein vector field and the scalars, do not take any particularly nice or universal form beyond what is dictated by symmetry.

\subsection{Supersymmetry and cases of interest \label{spci} }

Let us review the holographic interpretation of the backgrounds that we have just described. When $T_\pm=0$ and the $\l_i$ are small, they correspond to infinitesimal irrelevant deformations of $AdS_3\times S^3$. These deformations have been analysed in \cite{kerrdip,Kraus:2011pf}, using the catalogue in \cite{dkss}. All the Schr\"{o}dinger backgrounds \eqref{dipsch}, \eqref{genschm} are dual to deformations of the D1-D5 CFT by a $(1,2)$ operator. Even when the parameters $\l_i$ are not small, the interpretation as deformations by the corresponding irrelevant operators is conjectured to still hold, since these operators can be alternatively interpreted as exactly marginal  operators with respect to the non-relativistic conformal group $SL(2,\mathbb{R}) \times U(1)$ \cite{nr}.

 Each such operator is associated to a particular self-dual or anti-self-dual three-form field strength in six dimensions, present in the compactification of type IIB supergravity on $K3$ or $T^4$ \cite{Townsend:1983xt}. For example, the dipole Schr\"{o}dinger background is dual to defoming the D1-D5 CFT action by

\be
S_{dip} = S_{CFT} + \l \int du dv \, \O_{(1,2)}^{H^-} \label{defsdip}
\ee
where $H^-$ stands for the anti-self-dual part of the NS-NS three-form field. The $(1,2)$ operator above is a supersymmetric descendant of a $(1,1)$ chiral primary and preserves $(0,4)$ Poincar\'e supersymmetry \cite{Bobev:2011qx,kerrdip}. It has been explicitly checked in \cite{kerrdip} that the corresponding Schr\"{o}dinger background carries this amount of supersymmetry. 

In the case of the Ramond-Ramond backgrounds \eqref{genschm}, both the self-dual and anti self-dual parts of the RR three-form field turned on. Thus, they presumably correspond to the following type of deformations of the D1-D5 CFT

\be
S_{RR} = S_{CFT} + (\l_1+\l_2) \int du dv \, \O_{(1,2)}^{F^-} + (\l_1-\l_2) \int du dv \, \O_{(1,2)}^{F^+} \label{defsrr}
\ee
where we have used the fact that when $\l_1 = - \l_2$, the backgrounds \eqref{genschm} have purely self-dual three-form flux, whereas when $\l_1 = \l_2$ the  flux due to the deformation is purely anti-self-dual. The above two operators are again supersymmetric descendants of $(1,1)$ chiral primaries, preserving $(0,4)$ supersymmetry. Thus, we expect that all the backgrounds \eqref{genschm} are supersymmetric; this fact has been confirmed explicitly for the $\l_2=-\l_1$ background, for which the Killing spinors were explicitly constructed in \cite{Bobev:2011qx,kerrdip}.



Let us now turn to the black hole backgrounds. They are conjecturally dual to finite-temperature states in the theories defined by \eqref{defsdip} and \eqref{defsrr}. Note that since we only understand the holographic dictionary at zero temperature, it is not possible to determine whether the $\l_i$  appearing in the operator deformations above are exactly the same as the $\l_i$ appearing in \eqref{genbhm}; the only thing we know is that the two agree as $T_+ \r 0$. Thus, we could allow in principle for a relationship of the form

\be
\l_i^{\O} = \l_i \, f(\l_k T_+)
\ee
where $\l_i^\O$ are the couplings that appear in \eqref{defsrr} and $f$ is an arbitrary function with $f(0)=1$. The finite-temperature holographic dictionary we plan to develop should shed light on this issue. Note also that as soon as $T_+ \neq 0$, all the supersymmetry of these backgrounds is broken.

 The background with $\l_2=-\l_1$  has $F=\star F$ and appears in the string theory realization  \cite{microkerr} of the Kerr/CFT correspondence, reason for which we will call it ``6d NHEK'' or, even simpler, ``NHEK'', hoping that this terminology will not cause too bad of a confusion\footnote{Whenever necessary, we will refer to the original, four dimensional,  ``Near-Horizon-Extreme-Kerr'' geometry \cite{Bardeen:1999px} as $4d$ NHEK. }.  It appears in the six-dimensional uplift of the near-horizon geometry of a family of five-dimensional Kerr-Newman black holes. The background with $\l_1=\l_2$ is related by an $SO(n) \subset SO(5,n)$ duality rotation to the dipole background \eqref{dipbh} whose only effect is to turn the anti-self-dual NS-NS  field into the anti-self-dual part of the RR field.  We henceforth call these backgrounds ``S-dual dipole''.
Below we reiterate the data of the S-dual dipole and NHEK backgrounds in some detail, in order to have easy access to the necessary formulae later.

\bigskip

\noindent {\bf\emph{Dipole S-duals}}

\medskip

\noindent These backgrounds are obtained when the KK vector field vanishes, which from \eqref{akk} can be seen to happen when $\l_1=\l_2 =\l$. Also, the off-diagonal part of $F$ is anti self-dual. We then have

\be
\S_1 = \S_2 =\S\;, \;\;\;\;\;\Delta = (1-\l^2 T_+^2)^2\;, \;\;\;\;\tilde \l= \frac{2 \l}{1-\l^2 T_+^2} \label{tldip}
\ee

\be
ds^2  =  \ell^2 \sqrt{1+ \tilde \l^2 T_+^2}\left[ \frac{ds_3^2}{1+ \tilde \l^2 T_+^2}   +\frac{1}{4}\, d \Omega_2^2 + \frac{1}{4(1+ \tilde \l^2 T_+^2)} (d\psi+\cos\th d\phi)^2 \right] + \sum_i dx_i^2 \label{dipsd}
\ee
where $ds_3^2$ is given by \eqref{ref3dm}. Finally, 
\be
C^{(2)} = \frac{ \ell^2}{4} \cos \th d\phi \wedge d\psi - \ell^2\,  r d u \wedge d v + \frac{\tilde \l \ell^2}{2 \sqrt{1+ \tilde \l^2 T_+^2}} (r d u + T_+^2 d v) \wedge \s_3
\ee
The dilaton vanishes in this background.

\bigskip

\noindent {\bf\emph{NHEK backgrounds}}

\medskip

\noindent The condition that $F=\star F$ can be seen to be satisfied for $\l_1=-\l_2 = \l$. We have

\be
\Delta = \S^2 + 4  \l^2  T_+^2\;, \;\;\;\;\; \tilde \l = \frac{2 \l  \sqrt{3 + 2 \l^2 T_+^2 + 3 \l^4 T_+^4}}{(1-\l^2 T_+^2)^2}
\label{tlnhek} \ee
The KK gauge field is

\be
A= - \frac{4 \ell \l \S}{\Delta} \left( r du + T_+^2 d v \right)
\ee
and
\be
C^{(2)} = \frac{\ell^2}{4}  \cos \th d\phi \wedge d\psi - \ell^2\,  r du \wedge d v  + \frac{\l  \ell^2}{\S} \,\left( r du + T_+^2 dv \right)\wedge \s_3 
\ee
The metric bears no significant simplification over \eqref{genbhm} and the dilaton vanishes.

\section{Physical properties of the warped black strings \label{physprop}}

In this section we study the thermodynamic properties of the black string solutions that we have just found - entropy, temperatures, conserved charges - and show that they share certain universal features, similar to those of the BTZ black string. We  check that the first law of black hole mechanics holds and  briefly discuss asymptotic symmetries. Our analysis and results highly resemble those of the usual dipole black strings performed in \cite{stromwei}.


\subsection{Thermodynamics of the dipoles and NHEK}\label{DipNHEK}

In this subsection we analyze the thermodynamic properties of the six-dimensional S-dual dipole and NHEK black string
solutions, which are given by  (\ref{genbhm}) with $\l_1 = \pm \l_2$. As in \cite{stromwei},  we let 
\be
u=x+\tau \;, \;\;\;\;\; v = x -\tau
\ee 
and compute all charges and entropy per unit $x$-length. The generator of the horizon is taken to be 
\be
\xi_H = \p_\tau + \Omega_H \p_x \label{horgen}
\ee
from which we find the angular potential

\be
\Om_H = \frac{T_+-T_-}{T_++T_-}
\ee
The Hawking temperature is 

\be
T_{H} = \frac{2}{\pi} \frac{T_+ T_-}{T_+ + T_-}
\ee
Note the remarkable resemblance with the corresponding  thermodynamic potentials of  BTZ. This resemblance is due to the fact that the generalized spectral flows do not affect the thermodynamic properties of the horizon. The entropy per unit $x$-length is given by 

\be\label{entropy}
 S= \frac{\pi^2 \ell^4}{2 G_6} \, (T_+ + T_-)
\ee
We would now like to check whether the first law of black hole thermodynamics holds, namely that
\be
T_H \d S = \d E + \Om_H \d P \label{fstlaw}
\ee
where $E$ is the energy and $P$ the momentum of the black string  per unit $x$-length.

These conserved charges are easily computed using the Lagrangian formalism of \cite{bbch}. In this formalism, the difference in a particular conserved charge $Q_\xi$ - associated with the Killing vector/ asymptotic symmetry generator $\xi^\mu$ - between two backgrounds with metric $g$ and respectively $g + \d g$ is given by

\be
\d Q_\xi[\d g, g] =  \int_{\p \Sigma} K_\xi[\d g, g] \label{diffch}
\ee
The $(d-2)$ -form $K_\xi$ is constructed in a straightforward manner from the Lagrangian of the theory, and in general depends on both the metric and the matter fields and their infinitesimal variation. The integral is performed over a spacelike slice at constant radial distance, near infinity.

Upon removing the trivial $T^4$ factor, the black strings \eqref{genbhm} can be viewed as solutions of the following six-dimensional theory

\be
 S_{6d} = \frac{1}{16 \pi G_6}  \int d^6 x  \, \sqrt{g} \,\left( R - (\p \phi)^2 - \frac{1}{12} e^{2\phi} F^2\right)
\ee
The four-form $K_\xi$ can be written as

\be
K_\xi = \frac{1}{2\cdot 4!} \, \e_{\mu\nu\l\a\b\g} \, K_\xi^{\mu\nu} dx^\l \wedge dx^\a \wedge dx^\b \wedge dx^\g
\ee
where $K^{\mu\nu}_\xi$ is antisymmetric in its indices. For this theory, $K_\xi^{\mu\nu}$ receives three contributions: from the metric, the two-form field and the dilaton

\be
  K^{\mu\nu}_\xi = K^{\mu\nu}_{\xi,g} + K^{\mu\nu}_{\xi,C} + K^{\mu\nu}_{\xi,\phi} 
\ee
The expression for  $K_{\xi,g}$ is well-known 
\be
K^{\mu\nu}_{\xi,g} =\frac{1}{8\pi G_6} \left( \xi^\nu \nabla^\mu h - \xi^\nu \nabla_\s h^{\mu\s} + \xi_\s \nabla^\nu h^{\mu\s} + \half h \nabla^\nu \xi^\mu - h^{\rho \nu} \nabla_\rho \xi^\mu + \half h^{\s\nu} (\nabla^\mu \xi_\s + \nabla_\s \xi^\mu)\right)
\ee
and the one for $K_{\xi,C}^{\mu\nu}$ can be derived using  \cite{comp09,Compere:2007vx}\footnote{The formula below does not contain the contribution of the contracting homotopy \cite{Compere:2007az}. Nevertheless, this term never contributes to charges associated with exact Killing vectors of the background.} 
\bea
 K^{\mu\nu}_{\xi, C} &=&\frac{1}{16\pi G_6} \left( - 2 e^{2\phi} F^{\mu\nu\l} \xi^\rho C_{\rho \l} \, \d\phi - e^{2\phi} \xi^\rho \d C_{\rho \l} F^{\mu\nu\l} - \half \xi^\mu e^{2\phi } F^{\nu\a\b} \d C_{\a\b} + \half \xi^\nu e^{2\phi} F^{\mu\a\b} \d C_{\a\b} \right.  \non \\
&& +\, e^{2\phi} \xi^\rho C_{\rho \l} (h^{\mu\a} F_\a{}^{\nu\l} + h^{\nu\a} F^\mu{}_\a{}^\l 
 \left. + h^{\l\a} F^{\mu\nu}{}_\a  - \d F^{\mu\nu\l} ) - \half \, h \, e^{2\phi} F^{\mu\nu\l} \xi^\rho C_{\rho \l} \right)
\eea
where $\d F^{\mu\nu\rho} = g^{\mu\a} g^{\nu\b} g^{\rho\g} \d F_{\a\b\g}$. Finally,

\be
K^{\mu\nu}_{\xi,\phi} = \frac{1}{8 \pi G_6} (\xi^\nu \nabla^\mu \phi - \xi^\mu \nabla^\nu \phi) \, \d \phi
\ee
The energy and momentum of the string are given by the conserved charges associated to the Killing vectors $\p_\tau$ and $\p_x$
of the solutions. Since \eqref{diffch} only provides an expression for the \emph{difference} in energy/momentum between  two  backgrounds with neighbouring  metrics, we first need to compute $\d E$ and $\d P$ for two backgrounds whose temperatures differ by $\d T_\pm$, and then integrate with respect to the temperatures.  

Nevertheless, as we have previously discussed, it is not clear  \emph{a priori} whether the spectral flow parameters $\l_{1} = \pm \l_2 =\l$ are temperature-independent constants or not. Indeed, the parametrization in \eqref{tldip} and \eqref{tlnhek} suggests that one should perhaps keep $\tilde \l$, rather than $\l$, fixed as we vary $T_+$. We thus allow for a $T_+$ dependence of $\l_{1,2}$.
Interestingly, we find that the energy and momentum difference between two neighbouring backgrounds does not depend on $\l$ or its $T_+$-derivative at all 

\be
\d E = \frac{\pi \ell^4}{2 G_6} (T_+ \d T_+ + T_- \d T_-) \;, \;\;\;\;\; \d P =  \frac{\pi \ell^4}{2 G_6} (-T_+ \d T_+ + T_- \d T_-) 
\ee
Plugging these expressions into the first law of black hole thermodynamics \eqref{fstlaw}, we find perfect agreement. This agreement should not be surprising in view of the fact that each of the quantities entering \eqref{fstlaw} is identical between the BTZ black string and the NHEK/ dipole ones.  The above expressions can be trivially integrated to obtain the energy and momentum of the black strings, measured with respect to the Schr\"{o}dinger vacuum. 

\be
E=\frac{\pi \ell^4}{4 G_6} (T_+^2 + T_-^2) \;, \;\;\;\;\;  P =  \frac{\pi \ell^4}{4 G_6} ( T_-^2-T_+^2) 
\ee

\subsection{Thermodynamics of the general solutions}

The same analysis can be easily repeated for the case of the general black strings, \eqref{genbhm}, with $\l_1 \neq \pm \l_2$. The Hawking temperature and angular potential are the same as before

\be
T_H = \frac{2}{\pi} \, \frac{T_+ T_-}{T_+ + T_-} \;, \;\;\;\;\; \Om_H = \frac{T_+ - T_-}{T_+ + T_-}
\ee 
The entropy per unit length is now given by

\be
  S = \frac{\pi^2 \ell^4}{2 G_6} \, \frac{1 - \lambda_1^2 T_+^2}{1 - \lambda_2^2 T_+^2}\, (T_+ + T_-)
\label{uglyent}
\ee
The integrated energy and momentum per unit length are given by 

\be
E = \frac{ \pi \ell^4}{4 G_6}\, (T_+^2 +T_-^2) \, \frac{1-\l_1^2 T_+^2}{1-\l_2^2 T_+^2}\;, \;\;\;\;\; P=  \frac{ \pi \ell^4}{4 G_6}\, (T_-^2-T_+^2) \, \frac{1-\l_1^2 T_+^2}{1-\l_2^2 T_+^2}
\ee
We have adjusted the zero-point integration constants by demanding that $E=P=0$ when $T_\pm=0$. 
The electric and magnetic charges of the various backgrounds \eqref{genbhm} read

\be
Q_m = \frac{1}{4\pi^2} \int_{S^3} F = \ell^2 \;, \;\;\;\;\;\;\; Q_e = \frac{1}{4\pi^2}  \int_{S^3} e^{2\phi} \star F = \ell^2 \, \frac{1 - \lambda_1^2 T_+^2}{1 - \lambda_2^2 T_+^2}
\ee
Note that when $\l_1 \neq \pm \l_2$, varying $T_+$ with $\ell, \l_1,\l_2$ kept fixed amounts to varying the electric charge of the backgrounds we compare. This must be taken into account in the first law, which now reads

\be
T_H \d S = \d E + \Om_H \d P + \Phi_e \d Q_e 
\ee
As shown in \cite{Compere:2007vx}, the last term equals the integral over the horizon $r=T_+ T_-$ of the charge density associated to the horizon generator $\xi_H$, given in  \eqref{horgen}

\be
\Phi_e \d Q_e =  \int_{horizon} K_{\xi_H}  =- \frac{\pi \ell^4 (\l_1^2-\l_2^2) T_- T_+^2 \d T_+}{G_6(1-\l_2^2 T_+^2)^2}
\ee
With it, one can easily check that the first law is satisfied. The electric potential then reads 

\be
\Phi_e = \frac{\pi \ell^2}{2 G_6} T_+ T_-
\ee
Although the first law is satisfied as such, the expressions for the conserved charges and entropy would be extremely simplified if instead of $\ell$ we fixed the combination\footnote{Note that $\hat \ell/\ell$ can become imaginary when $|\l_1 T_+| < 1 < |\l_2 T_+|$ and vice-versa. We have no intuition for the significance of this fact.}

\be
\hat \ell^4 \equiv  \,  \frac{1 - \lambda_1^2 T_+^2}{1 - \lambda_2^2 T_+^2}\, \ell^4 = Q_e Q_m \label{deflhat}
\ee 
In terms of $\hat \ell$, the entropy and conserved charges take the remarkably simple form

\be
S = \frac{\pi^2 \hat \ell^4}{2 G_6} (T_+ + T_-) \label{newentf}
\ee

\be
E=\frac{\pi \hat \ell^4}{4 G_6} (T_+^2 + T_-^2) \;, \;\;\;\;\;  P =  \frac{\pi \hat \ell^4}{4 G_6} ( T_-^2-T_+^2) 
\ee
which is very reminiscent of the Cardy formula in a CFT. Note that $\hat \ell^4$ is proportional to the product of the electric and the magnetic charge of the system. Thus, fixing $\hat\ell$ corresponds in a certain sense to fixing the ``central charge'' of the associated field theory dual, which for systems similar to the D1-D5 black string equals the product of the electric and the magnetic charges. That $\hat \ell^4$ corresponds to a central charge will become clear in the asymptotic symmetry group analysis of the next subsection.

Given the simplicity of the above expressions, one may wonder whether we should have fixed $\hat \ell$ rather than $\ell$ from  the very beginning of the thermodynamic analysis. Nevertheless, in this case we find that the energy difference between two backgrounds  is divergent as $r \r \infty$ whenever $\l_1 \neq \pm \l_2$

\be
\d E =  \frac{ \pi \hat \ell^4}{2G_6} \, \frac{ (\l_1^2-\l_2^2) T_+ \d T_+}{(1-\l_1^2 T_+^2)(1-\l_2^2 T_+^2)} \, r +  \frac{ \pi \hat \ell^4}{2G_6} (T_+ \d T_+ + T_- \d T_-)
\ee
That we find a linear divergence with the radius, or even any radial dependence at all, is extremely surprising, since charges associated to exact Killing vectors should not depend on the surface $\p \S$ on which we choose to evaluate them. Nevertheless,  this divergence can be understood already in pure $AdS_3 \times S^3$: if we compare two backgrounds of different radii, $\ell$
and $\ell + \d \ell$, we will find an infinite energy difference, due to the fact that the two backgrounds have different asymptotics (one has to change the non-normalizable part of the metric in order to get from one to the other).
Thus, if we do not want to have infinite energy, we should better keep the magnetic charge fixed as we vary the temperatures. This can be achieved either by fixing $\ell$ instead of $\hat \ell$, or by fixing $\hat \ell$ and $\l_i T_+$ as we vary $T_+.$



It is very interesting that all the above black holes have a thermodynamics that is completely independent of the deformation parameters $\l_i$ - in an appropriate parametrization -  and resembles significantly that of the BTZ black hole. In the following subsection we will reinforce this observation by computing the asymptotic symmetry data.
%



\subsection{Asymptotic symmetry group analyses}

As is usual in warped $AdS_3$, there are various choices of boundary conditions that lead to a well-defined asymptotic symmetry algebra, where the charges are finite, integrable and conserved \cite{stromwei}. One such choice leads to an asymptotic symmetry group that enhances the $U(1)$ isometry of the spacetime to a Virasoro; another choice leads to a Virasoro enhancement of the $SL(2,\mathbb{R})$ isometries of the vacuum spacetime, 	and a simultaneous enhancement of the $U(1)$ isometry to a $U(1)$ Ka\v{c}-Moody current algebra. We will discuss each in turn.

\bigskip
\noindent {\bf \emph{Enhancement of U(1) to Virasoro}}

\medskip

\noindent
 As shown in \cite{kerrdip,stromwei} for the very similar dipole background, there exist consistent boundary conditions on the metric and gauge field fluctuations,  such that the $U(1)$ isometry is enhanced to a Virasoro asymptotic symmetry.
The relevant boundary conditions are spelled out in \cite{kerrdip}. The asymptotic symmetry group generators are associated to the vector fields

\be
\xi_f = f(v) \, \p_v - f'(v)\, r \p_r
\ee
Expanding $f(v)$ in Fourier modes, it can be easily shown that the Lie bracket algebra of the corresponding vector fields $\xi_n$ is a Virasoro algebra. The Dirac bracket algebra of the associated symmetry generators $Q_{\xi_n}$ is again a Virasoro algebra\footnote{The charges are finite and conserved only if we integrate the charge densities \eqref{diffch} along the $v$ direction. Note that this is a different prescription than we used in the two previous subsections.}, now with a non-trivial central extension

\be
c = \frac{ 3 \pi^2 \hat \ell^4}{G_6} 
\ee
where $\hat \ell$ is given by \eqref{deflhat}.  Choosing conventions in which $G_6 = \frac{\pi^2}{2}$, we recover the familiar expression
\be
c= 6 \hat \ell^4 = 6\, Q_e Q_m
\ee
Positing the existence of a second, left-moving Virasoro algebra with an identical central charge,
we can check that the entropy per unit length satisfies Cardy's formula

\be
S = \frac{\pi}{6} \, c \, (T_L + T_R) 
\ee
where $T_L = \pi^{-1} T_-$, $T_R = \pi^{-1} T_+$. Understanding the origin of this Virasoro symmetry and the reason that the entropy is captured by Cardy's formula is an important question that we hope future holographic analyses will answer. Note that we can only meaningfully apply Cardy's formula if the central charge is a temperature-independent constant, which suggests that we should keep $\hat \ell$, rather than $\ell$ fixed as we compare the various backgrounds.


\bigskip
\noindent {\bf \emph{Enhancement of U(1) to Ka\v{c}-Moody}}

\medskip

\noindent
There is another set of consistent boundary conditions, first discussed in \cite{Compere:2007in}, that have been recently used to study the dipole backgrounds \cite{Detournay:2012pc}. The asymptotic symmetries  
in this case consist of the semi-direct product of a Virasoro and a Kac-Moody algebra,
whose generators can be taken to be

\be
   L_n =   e^{i n u} \left( \p_u -  i n r   \p_r \right)\;, \;\;\;\;\;
   P_n =  -  e^{i n u}  \p_v  
\ee
The asymptotic symmetry algebra reads

\be
[L_m,L_n]= (m-n) L_{m+n} + \frac{c}{12} (m^3-m) \d_{m+n} 
\ee

\be
[L_m,P_n] = - n P_{m+n} + n P_0 \d_{m+n} \;, \;\;\;\;\;\;
[P_m,P_n] = 2 m P_0  \d_{m+n}
\ee
Notice that the algebra takes a slightly unorthodox form,  the $\widehat{u(1)}$ level depending on the zero mode charge $P_0$.  The asymptotic density of states in a theory described by this algebra is given by \cite{Detournay:2012pc}

\be \label{SWCFT}
  S = 4 \pi \sqrt{- P_0^{vac}  P_0} + 4 \pi \sqrt{- L_0^{vac}  L_0}
\ee
where $L_0^{vac}$ and $P_0^{vac}$ are the corresponding charges computed in the vacuum state. The left-moving energy $L_0^{vac}$ of the vacuum on the cylinder is easy to compute, by the usual conformal mapping to the plane, yielding

\be \label{L0Vac}
  L_0^{vac} = -\frac{c}{24}
\ee
Nevertheless, $P_0^{vac}$ is a parameter of the theory that is not fixed by the algebra or the minimum energy condition, and we need to fix it in a different way, discussed below. 

To compute the charges, we integrate the symmetry generators along the $x$ direction and we find that they are finite. The corresponding zero mode charges are 

\be
L_0 = \frac{\pi \hat \ell^4  T_-^2}{4 G_6}\;, \quad \quad P_0 = \frac{\pi \hat \ell^4 T_+^2}{4 G_6}
\ee
whereas the central extensions read
\be\label{MeLabelYou}
 c = \frac{3 \pi \hat \ell^4}{2 G_6}\;, \quad \quad k = -\frac{\pi \hat \ell^4 T_+^2}{G_6} 
\ee
Note that even in the case $\l_1 = \pm \l_2$ when $\hat \ell$ is constant, the level depends on the state, and is generally negative. The energy of the vacuum is \eqref{L0Vac}, and can be formally obtained by setting
\be\label{TmVac}
  T_-^{vac}  = \pm \frac{i}{2}
\ee
Since only $T_-^2$ appears in the metric, this value does not yield imaginary fields. It is easy to see that with this value, the second term in \eqref{SWCFT} precisely matches the second term in the Bekenstein-Hawking entropy formula \eqref{newentf}.

Matching the right-moving contribution to the entropy is somewhat tricky. The value of $P_0$ in the vacuum that we would need is
\be
P_0^{vac} =  -\frac{\pi \hat \ell^4}{16 G_6} 
\ee
Again, this value formally leads to an imaginary right-moving temperature

\be\label{caramelmou}
  T_+^{vac} = \pm \frac{i}{2}.
 \ee
When the deformation is set to zero, $\l_1=\l_2=0$, this value is easy to argue for: it is precisely the value for which the undeformed BTZ metric becomes global\footnote{ It is well known that in BTZ this unique value can be obtained by requiring that the negative-mass BTZ solution be free of conical defect singularities when $x$ is compact. The same logic can be applied here, as one can easily show that the metric \eqref{genbhm} with $x$ compact is  free of conical defects for imaginary values of $T_\pm$ only when \eqref{TmVac} and \eqref{caramelmou} are satisfied.} $AdS_3$, provided we also use \eqref{TmVac}. Nevertheless, as soon as we turn on the deformation, the vector fields will all become imaginary for this value of $T_+$, and it is hard to make sense of the solution.

In conclusion, while the left-moving contribution to the entropy easily matches for all values of $T_\pm$, the right-moving part only matches if we are willing to allow imaginary gauge fields. Moreover, for all backgrounds with $\l_1 \neq \pm \l_2$,
we are forced to keep $\hat \ell$ rather than $\ell$ fixed as we compare them.
Thus, the warped CFT symmetries are able to reproduce the entropy of the warped black strings, as was already observed for the dipole geometries in \cite{Detournay:2012pc}, but only subject to the above rather big concessions.

\section{Consistent truncations to three dimensions \label{seccttr}}

One of the main goals of this paper is to find simplified three-dimensional models that capture the most relevant physics of the six-dimensional backgrounds described in section \ref{nwbh}. 
We are particularly interested in knowing whether the three-dimensional parts of the metrics \eqref{genbhm} fit inside  consistent truncations of type IIB string theory to three dimensions, thus providing warped  analogues of the BTZ black hole in three dimensions. 

These consistent truncations will generically contain three-dimensional gravity, massive vectors and scalars.
A consistent truncation for the three-dimensional dipole backgrounds can be easily found following the steps of \cite{maldanr}, who studied the analogous five-dimensional case. We could also easily find a consistent truncation with only Ramond-Ramond flux that contains all the Schr\"{o}dinger backgrounds \eqref{genschm}. Nevertheless, the general warped black string backgrounds \eqref{genbhm} fit inside this truncation only if $\l_1 = \pm \l_2$, that is only for the S-dual dipole and NHEK backgrounds. We show that in both these cases the three-dimensional action can be truncated to even fewer fields, while still capturing the black hole solutions of interest.

In the following subsections we will present four consistent truncations: first, for the usual dipole background, which is a simple adaptation of \cite{maldanr} to three dimensions; second, for the general Schr\"{o}dinger backgrounds supported by Ramond-Ramond flux. Next, we further truncate the latter action to as few fields as possible, while maintaining the requirement that the truncation be consistent. The details of all the consistent truncations can be found in appendix \ref{cttrunc}.


\subsection{Consistent truncation with NS-NS and RR flux}

We start from the consistent truncation of type IIB supergravity  to six dimensions  described in  \cite{duffm}, containing the metric, the NS-NS $H$-field, the Ramond-Ramond three-form $F$-field and two axion-dilatons
\bea
 S_{6d} & = & \frac{1}{16 \pi G_6} \int d^6 x \, \sqrt{g} \, \left( R - \half \sum_{i=1}^2 \left[ (\p \phi_i)^2 + (\p \chi_i)^2 e^{2 \phi_i} \right] - 
\frac{1}{12} e^{-\phi_1-\phi_2} H^2 -\right. \non \\  && \hspace{6cm} \left. - \frac{1}{12} e^{\phi_1-\phi_2} F^2 + \chi_2 \, H \wedge F \right) \label{6dact}
\eea
with $H=dB$ and  $F = d C^{(2)} + \chi_1 H$. Here $\phi_1$ is the ten-dimensional dilaton (previously denoted as $\Phi$), $e^{-\phi_2}$ is the  volume of $K3$ or $T^4$ in the ten-dimensional Einstein frame, $\chi_1$ is the ten-dimensional axion and
$\chi_2$ is the six-dimensional dual of $F_5$. In order to describe  the dipole backgrounds, which have $F=\star F$ and $F \wedge H =0$, we would like to further truncate this action to such configurations. It is consistent to concomitently fix some of the scalars to
\be
\chi_1 =0 \;, \;\;\;\;\; \chi_2 = const \;, \;\;\;\;\; 
\phi_1 = \phi_2 \equiv \phi
\ee 
The field $\phi$ above equals the six-dimensional dilaton. The resulting six-dimensional action reads

\be
S_{6d} = \frac{1}{16\pi G_6} \int d^6 x \, \sqrt{g}\, \left( R - (\p \phi)^2 - \frac{1}{12} e^{-2\phi} H^2 - \frac{1}{12} F^2 + \chi_2\, H \wedge F \right) \label{truncact}
\ee
Note this truncation of the IIB action resembles very much the ten-dimensional Ansatz in \cite{maldanr}. 
Let us now make the following Ansatz for reducing to three dimensions on a squashed three-sphere

\be
ds_6^2 = e^{-4 U - 2 V} ds_3^2 + \frac{\ell^2}{4} e^{2 U} d\Omega_2^2 + \frac{\ell^2}{4} e^{2 V} (d\psi + \cos \th d \phi)^2 
\ee

\be
B =\frac{\ell}{2} \, \hat A \wedge (d\psi + \cos\th d\phi ) \;, \;\;\;\;\; F = (1+ \star_6) f_3 \;, \;\;\;\;\; f_3 = \frac{2}{ \ell} e^{-4 (2 U + V)} \,\omega_3
\ee
where $\omega_3$ is the volume form associated to the three-dimensional metric $ds_3^2$.
The truncation to a three-dimensional model is consistent if the six-dimensional equations of motion resulting from the action \eqref{truncact} are automatically solved by all solutions to the three-dimensional equations of motion. That this is the case is proven in the appendix \ref{ctudip}. Moreover, we can make a further truncation to solutions that have

\be
V = - U
\ee
The three-dimensional equations of motion are written in the same appendix. They  can be derived from the following effective  action 
\bea
S_{3d} & = &  \int d^3x \sqrt{g} \left( {}^{(3)} R - 4 ( \p U)^2 - ( \p \phi)^2  + \frac{4}{\ell^2}  e^{-4 U} - \frac{2}{ \ell^2} e^{-8 U}- \right. \non \\
&& \hspace{6 cm} \left.- \frac{2}{ \ell^2} e^{-2 \phi -4 U} \hat A^2 - \frac{1}{4} e^{-2 \phi + 4 U} \hat F^2 \right) \label{dipact}
\eea
Consequenty, we have succeeded in describing the relevant part of the low-lying sector of type IIB supergravity around dipole backgrounds by just three fields: two scalars, $\phi$ and $U$ and one massive vector field, which carries two degrees of freedom.

\bigskip
\noindent {\bf \emph{The three-dimensional dipole black hole}}

\medskip

\noindent We can use the above reduction to write the dipole black string solution \eqref{dipbs} in three-dimensional language. Remembering that the $6d$ dilaton is $\phi$, passing to six-dimensional Einstein frame and then  using our consistent truncation Ansatz, we  find that

\be
e^{4U} =  e^{-2\phi} = 1 + \l^2 T_+^2
\ee
Consequently, the three-dimensional Einstein metric we obtain is

\be
ds_3^2 = \ell^2 \left( T_+^2 dv^2 + 2 r \, du \, dv + \left[ T_-^2 (1+ \l^2 T_+^2) - \l^2 r^2\right] du^2 + \frac{ dr^2}{4 (r^2 - T_+^2 T_-^2)} (1+ \l^2 T_+^2) \right) \label{3deinst}
\ee
and massive vector field
\be
\hat A =  \frac{\l \ell}{1+ \l^2 T_+^2} (T_+^2 dv + r du)
\ee
One can check that these fields satisfy  the three-dimensional equations of motion \eqref{mxappb}.

\subsection{Consistent truncations with only RR flux}

Let us now concentrate on the consistent truncations for the backgrounds supported by purely RR flux. Since $H=0$, the six-dimensional equations of motion derived from \eqref{6dact} are consistent with the choice

\be
\phi_1 = - \phi_2 = \phi\;, \;\;\;\;\; \chi_1=\chi_2 =0
\ee
The six-dimensional dilaton vanishes. The six-dimensional action becomes

\be
 S_{6d} = \frac{1}{16\pi G_6} \int d^6 x \sqrt{g} \, \left( R -(\p\phi)^2 - \frac{1}{12} e^{2\phi}  F^2 \right) \label{pureRR}
\ee
We consider the following Ansatz for the reduction from six to three dimensions

\be
ds_6^2 = e^{-4 U - 2 V} ds_3^2 + \frac{\ell^2}{4} e^{2 U} d\Omega_2^2 + \frac{\ell^2}{4} e^{2 V} \left(d\psi + \cos \th d \phi + \frac{2}{\ell} \, A\right)^2 \non
\ee

\be\,
F = \frac{2}{\ell} \, X \, \om_3+ \frac{\ell^2}{4} d\Omega_3 + \frac{\ell}{2}\, d \hat C \;, \;\;\;\;\; \hat C = \hat  A \wedge \s_3 \label{genansatz}
\ee
where

\be
d\Omega_2 = \sin\th \, d\th \wedge d\phi \;, \;\;\;\;\; \s_3 = d\psi + \cos \th \, d\phi \;, \;\;\;\;\; d\Omega_3 = d \Omega_2 \wedge \s_3
\ee
and $\om_3$ is the volume form associated with the $3d$ Einstein metric $ds_3^2$. Note that $dF=0$ by construction. The Maxwell equation $d(e^{2\phi} \star F) =0$ requires that the coefficient in front of $d\Om_3$ in the expression for $e^{2\phi} \star F$ be a constant, which we fix  by  requiring that $AdS_3$ of radius $\ell$ be a solution. With this 

\be
X = e^{-8U-4V-2\phi} + \frac{\ell}{4} {}^{(3)} \e^{mnp} \hat F_{mn} A_p
\ee
Note that the general black hole solutions \eqref{genbhm} don't satisfy the above  requirement, except when $\l_1 = \pm \l_2$, since they have 

\be
e^{2\phi}\star F = \frac{\ell^2}{4} \frac{1-\l_1^2 T_+^2}{1-\l_2^2 T_+^2} \, d\Om_3 + \ldots
\ee
and thus will not satisfy the Maxwell equation once we promote the  $T_+$-dependent term  to a $x^\mu$-dependent function. Thus, the black string solutions with $\l_1 \neq \pm \l_2$ will not fit inside the consistent truncation we have found. It would be interesting to find  a different truncation that  accomodates them, but we were unable to. 

The six-dimensional equations of motion imply a set of equations for the three-dimensional fields which can be derived from the following $3d$ action

\bea
S_{3d} &=& \int d^3x \sqrt{g} \left[{}^{(3)}R - 6 (\p U)^2 -2(\p V)^2 - 4 \p_i U \p^i V - (\p \phi)^2 - \frac{1}{4} \,e^{4U+4V} \,F^2 - \frac{1}{4} \, e^{4U+2\phi} \hat F^2  \right. \non \\
&&\hspace{-1cm} \left. - \frac{2}{\ell^2} \, e^{-4U+2\phi} (\hat A-A)^2 + \frac{2}{\ell^2} \, e^{-8U}  \left(4\, e^{2U-2V} -1 - e^{2\phi-4V} -  e^{-2\phi-4V} \right) + \frac{1}{\ell}\, \e^{ijk} A_i \hat F_{jk}\right] \non \\ && \label{gentract}
\eea
Naively, one would think this action describes seven propagating degrees of freedom, but in fact there are only six, due to the relation \eqref{constrtwoas} between $A$, $\hat A$ and their field strengths, which holds up to the addition of a closed one-form.  

\subsection{Further consistent truncations}

As we have just seen, the action \eqref{gentract} can describe the finite-temperature black strings only in the special cases $\l_1 = \pm \l_2$. In both of these cases we are able to further  truncate the action to a smaller set of fields, so as to obtain a much simpler three-dimensional action which does contain black hole solutions. 

\subsubsection*{Consistent truncation for the S-dual dipoles}

For the S-dual dipoles, we note that the solution \eqref{dipsd} has vanishing  Kaluza-Klein vector field. Consequetly, we start from the Ansatz \eqref{genansatz}, but we set $A=0$. The six-dimensional equations of motion imply that

\be
X=e^{-8U -4V-2\phi} \;, \;\;\;\;\;\hat F_{ij} = \frac{2}{\ell} e^{2
\phi -4U} \e_{ijk} \hat A^k \label{newmax}
\ee
Moreover, we can safely truncate to backgrounds where

\be
\phi =0 \;, \;\;\;\;\; V = - U
\ee
The three-dimensional action from which all the $3d$ equations can be derived reads

\be
S= \int d^3 x \sqrt{g} \left( R - 4 (\p U)^2 - \frac{4}{\ell^2} \, e^{-4U} \, \hat A^2 + \frac{2}{\ell^2} \, e^{-4U} (2-e^{-4U}) -\frac{1}{\ell} \, \e^{ijk} A_i F_{jk}\right) \label{actsddip}
\ee
This is simplest truncation we were able to find that contains warped black hole solutions. It consists of one scalar, $U$, and one chiral massive vector field, which carries only one degree of freedom. The details of the truncation and the equations of motion can all be found in appendix \ref{cttrsddip}.

\bigskip

\noindent {\bf\emph{The S-dual dipole black hole}}

\medskip

\noindent In terms of the fields in this truncation, the S-dual dipole solution to the action \eqref{actsddip} has

\be
e^{2U} = \sqrt{1+\tilde{\l}^2 T_+^2}\;, \;\;\;\;\;\hat A= \frac{\tilde\l \ell}{\sqrt{1+\tilde \l^2 T_+^2}} (T_+^2 dv + r du)
\ee
The $3d$ Einstein metric takes the same form as for the usual dipoles

\be
ds_3^2 =\ell^2 \left( T_+^2 dv^2 + 2 r \, du \, dv + \left[ T_-^2 (1+ \tilde \l^2 T_+^2) - \tilde \l^2 r^2\right] du^2 + \frac{ dr^2}{4 (r^2 - T_+^2 T_-^2)} (1+ \tilde \l^2 T_+^2) \right)
\ee
with $\l$ replaced by $\tilde\l$. According to  \eqref{tldip}, the parameter $\tilde \l$ is some particular function of $\l$ and $T_+$. If we assume that it is the spectral flow parameter $\l$ that should be kept constant as we vary the temperature, then the above metric has a complicated $T_+$ dependence. Nevertheless, it does seem natural from a holographic point of view to instead fix $\tilde \l$ as $T_+$ is varied; the striking resemblance between the two dipole metrics suggests this. As we have shown in the previous section, the thermodynamics of the dipole black strings is independent of both $\l$ and $\tilde \l$, so it is unaffected by either choice. One may hope that a proper understanding of holography for these backgrounds will fix this ambiguity.

\subsubsection*{Consistent truncation for NHEK}

The  NHEK background $\l_1=-\l_2$ has $F=\star F$ and $\phi=0$, conditions which are compatibe with each other, since 

\be
\phi =0  \;\;\;\;\; \Rightarrow \;\;\;\; F^2 =0
\ee
for the action \eqref{pureRR}. The consistent truncation Ansatz  is again \eqref{genansatz}.  Imposing the self-duality condition $F=\star F$ we obtain

\be
\frac{2}{\ell} ( \hat A -  A) =  e^{4U} \star_3 \hat F 
\ee 
The equations of motion further imply that

\be
e^{4U+4V} \star_3 F= - \frac{4}{\ell} \hat A
\ee
The truncated three-dimensional action describing these fields is

\bea
S_{3d} &=& \int d^3x \sqrt{g} \left[{}^{(3)}R - 6 (\p U)^2 -2(\p V)^2 - 4 \p_i U \p^i V -\frac{ 8}{\ell^2} \, e^{-4U-4V} \hat A^2 - \frac{ 4}{\ell^2} \,  e^{-4U} (\hat A -A)^2 + \right. \non \\
&& \hspace{1cm} \left.  + \frac{2}{\ell^2} \, e^{-8U}  \left(4\, e^{2U-2V} -1 - 2 e^{-4V} \right) -\frac{2}{\ell}\, \e^{ijk} \left( A_i -\half \hat A_i \right) \hat F_{jk}\right] \label{nhekact}
\eea
Thus, the truncation that contains NHEK is slightly more involved, containing two scalars, $U$ and $V$ and two coupled chiral massive vector fields. It thus describes four propagating degrees of freedom in three dimensions.

\bigskip
\noindent {\bf\emph{The NHEK black hole}}

\medskip

\noindent The Einstein-frame metric takes the same form as before

\be
ds_3^2 =\ell^2 \left( T_+^2 d v^2 + 2  r d u dv + [T_-^2 (1+ \tilde \l^2 T_+^2) - \tilde \l^2  r^2] du^2  +\frac{ d r^2}{ 4( r^2 - T_+^2 T_-^2)} (1+ \tilde \l^2 T_+^2) \right) \label{nicemet}
\ee
The parameter $\l$ is a complicated function of $\l$ and $T_+$, given in \eqref{tlnhek}. Again, it is not clear whether it is $\l$, $\tilde \l$ or some other combination that we should fix as we vary the temperature. The  scalar fields read 

\be
e^{2U} = \frac{\S }{1-\l^2 T_+^2} \;, \;\;\;\;\; e^{2V} = \frac{\Delta}{\S (1-\l^2 T_+^2)}
\ee
where the expressions for $\Delta$ and $\S$ are given in \eqref{tlnhek} and \eqref{tldip}.  We were unable to find a simple way of writing these expressions as a function of $\tilde \l$, although it is easy to show that 

\be
e^{6U+2V} = 1+\tilde \l^2 T_+^2
\ee
The two chiral massive vector fields  read\footnote{The reason for the additional minus sign in $\hat A$ is that it is $-C^{(2)}$ in \eqref{genbhcf}, rather than $C^{(2)}$, that obeys the Ansatz \eqref{genansatz}.}

\be
A = -\frac{4 \l \S \ell}{\Delta}\, (rdu + T_+^2 dv) \;, \;\;\;\;\; \hat A = - \frac{2 \l \ell}{\S}\, (rdu + T_+^2 dv)
\ee

\section{The string theoretical spectra \label{strspec}}

For a free massive scalar field propagating in a Schr\"{o}dinger background, the fall-off of the solution at large distances is given by

\be
\phi \sim \phi_s \, r^{ h-1} + \phi_{vev} \, r^{- h } 
\ee
where

\be
h=\half + \sqrt{\frac{1}{4} + \l^2 \k^2}
\ee
is the $SL(2,\mathbb{R})_L$ conformal weight of the field, and  $\k$ represents the momentum along the null $v$ direction.
According to the nonrelativistic AdS/cold atom correspondence \cite{Son:2008ye,Balasubramanian:2008dm}, $2h$ represents the nonrelativistic conformal dimension of the operator dual to the scalar field.   Thus, it appears that the non-relativistic conformal dimension always increases with the null momentum. Is this a general feature of the perturbative spectrum about a Schr\"{o}dinger background, or $h$ can also decrease/ become imaginary as we increase the momentum? Imaginary $SL(2,\mathbb{R})$ conformal weights are particularly interesting in the study of the original (four-dimensional) Kerr/CFT correspondence, as it was shown \cite{Dias:2009ex} that the spectrum of gravitational perturbations around $4d$ NHEK always allows modes with such behaviour.

Given the consistent truncations that we have just described, we can use them to find part of the exact  spectrum of $SL(2,\mathbb{R})$ conformal dimensions of type IIB perturbations about the various three-dimensional Schr\"{o}dinger backgrounds \eqref{genschm} and answer the  question above. Quite surprisingly, we find that most backgrounds - including zero-temperature NHEK - contain modes with imaginary fall-offs. We briefly address the issue of classical linearized stability of the $6d$ NHEK background in section \ref{stabanal}.

\subsection{Setup and generalities about the spectrum}

In this section we study the perturbative spectrum of type IIB supergravity about the various Schr\"{o}dinger backgrounds \eqref{genschm} using the consistent truncations developed in the previous section. The three-dimensional background metric  takes the form

\be
d\bar s^2 = \ell^2\left(\frac{2  du dv}{\rho} - \frac{ \tilde{\l}^2  du^2}{\rho^2} + \frac{d\rho^2}{4\rho^2} \right) \label{genbckm}
\ee 
where $\tilde \l$ is given by \eqref{tl} and we have changed coordinates to $\rho = r^{-1}$. The background values of the massive vector fields are

\be
\bar{\hat{A}} = - \frac{2 \l_1 \ell}{\rho} du \;, \;\;\;\;\; \bar{A} = -\frac{2(\l_1-\l_2) \ell}{\rho} du \label{genbcka}
\ee
and all the background scalars are zero $\bar U = \bar V = \bar \phi =0$. To study the spectrum, we solve the linearized equations of motion derived from the action \eqref{gentract}\footnote{Of course, for the usual dipole backgrounds we need to use the action \eqref{dipact} and perturb around the corresponding background solution. } about these Schr\"{o}dinger backgrounds. Concretely, we let

\be
g_{\mu\nu} = \bar g_{\mu\nu} + \e \, h_{\mu\nu} \;, \;\;\;\;\; A_\mu^I = \bar A_\mu^I + \e \, \mathcal{A}_\mu^I\;, \;\;\;\;\; \phi^a = \bar \phi^a + \e \, \varphi^a\;, \;\;\;\;\; \e <<1
\ee
where $I$ labels the various massive vector fields and $a$ the various scalars. We work in radial gauge for the metric perturbation

\be
h_{i \rho} = h_{\rho\rho} =0\;, \;\;\;\;\;\;\;\; i \in \{u,v\}
\ee
Next, we solve the full coupled system of linearized equations of motion using Mathematica. Since the details of this computation are too complicated to be transcribed herein, in the following we just outline the main steps.

The qualitative features of the spectrum of linearized perturbations of a given theory around the three-dimensional Schr\"{o}dinger vacuum are well known. In radial gauge, one finds two types of modes, which in \cite{nr} have been denoted as $T$-modes and $X$-modes

\be
h_{ij} = h_{ij}^X + h_{ij}^T \;, \;\;\;\;\; \mathcal{A}_\mu = \mathcal{A}_\mu^X + \mathcal{A}_\mu^T\;, \;\;\;\;\; \varphi = \varphi^X
\ee
The so-called $T$-modes  have been extensively studied in \cite{fg}. Just like solutions of pure Einstein gravity in $AdS_3$, 
they  can be induced by radial-gauge-preserving diffeomorphisms. Thus, they are universal: they don't depend on the specific theory one is studying, as long as it admits a Schr\"{o}dinger solution.  They have integer or at most logarithmic power fall-offs at large radius, and they are in one-to-one correspondence with solutions of  pure Einstein gravity  in AdS$_3$ \cite{fg}. 

One the other hand, the $X$ modes correspond to non-trivial propagating degrees of freedom. They are best studied in Fourier space, so we let

\be
h_{ij}(u,v,\rho) = \frac{1}{2\pi} \int du dv \, \tilde h_{ij} (\om, \k ,\rho)\, e^{- i \om u + i \k v} 
\ee
and similarly for the $\mathcal{A}_\mu^I$ and $\varphi^a$. The $X$ modes have characteristic falloffs that depend on the momentum $\k$ of the perturbation along the null direction $v$, as well as on the specific couplings in the Lagrangian. 

In order to find the $X$-mode solutions corresponding to a given theory, one needs to solve the fully coupled system of linearized equations of motion about the vacuum Schr\"{o}dinger spacetime of interest, including all the massive vector fields and the scalars. The reason is that, unlike in asymptotically AdS spacetimes\footnote{In the case of deformations by sufficiently low-dimension relevant operators only. For counterexamples, see e.g. \cite{Donos:2011bh,Donos:2011qt}. }, the conformal dimensions of the various dual operators are extremely sensitive to the specific form of the couplings in the supergravity Lagrangian. 

The solution  is usually found using Mathematica. One expresses all the components of the metric, massive vectors and the various scalars in terms of a single, conveniently chosen scalar field or tensor component - call it $g_{\k, \om}(\rho)$. 
The various components of the metric, massive vectors and scalars are rather ugly-looking expressions involving $g_{\k,\om}$,  its radial derivatives and complicated functions of the momenta $\k,\om$.

 Since in general all the degrees of freedom in the action are coupled, the function $g_{\k, \om}(\rho)$  obeyes a $2n$-order linear differential equation, where $n$ is the number of propagating degrees of freedom in the system. Thus, we may schematically write


\be
\sum_{k=0}^{2n} a_k \, g^{(k)}(\rho) =0 \label{geneqng}
\ee
This equation has certain special properties. First, the coefficient functions $a_k$ are such that the only  radial dependence in this equation is through the variable $y \equiv \rho\, \om \k$. Secondly, the coefficient functions only depend on $y$ and the combinations $(\l_i \k)^2$. Thus,

\be
g=g(y) \;, \;\;\;\;\; a_{k} = a_k(y, \l_i^2 \k^2 ) \;, \;\;\;\;\;y= \rho\, \om \k
\ee
The equation \eqref{geneqng} has a regular singular point at $y=0$ and usually an irregular singular point at $y \r \infty$. The exact solution has been obtained only for $n=1$ (as in the TMG example \cite{nr}) and also - with a lot of work - for $n=2$ in the massive vector case \cite{vanRees:2012cw}. Nevertheless, for the purpose of knowing the spectrum of operators of the dual theory we only need to understand the behaviour near $y=0$. Letting

\be
g(y) = y^s + \O(y^{s+1}) 
\ee
then to leading order in $y$ the differential equation obeyed by $g$ turns into a degree $2n$ polynomial equation for $s$, with roots $s_k$. Interestingly, the roots always come in pairs that satisfy

\be
s_{k} + s_{k+1} =1 \;, \;\;\;\;\; k \;\,\mbox{odd} \label{wtcons}
\ee
The numbers $s_k$ characterize the radial falloff the $X$-mode perturbations. As is usual in holography, the radial falloff in spacetime is related to the conformal dimension of the holographically dual operator\footnote{This is true only if the perturbation transforms in a highest weight representation of the conformal group - in this case $SL(2,\mathbb{R})$.}. In the following we will find the spectrum $s_k$  of  $SL(2,\mathbb{R})_L$  conformal dimensions of  the perturbations around the $Schr_3$ background in several of our consistent truncations, and discuss its features. 

\subsection{Features of the spectra}

As we have just explained, the spectrum of perturbations inside a particular consistent truncation depends on the number of degrees of freedom kept in the lower-dimensional Lagrangian: the fewer, the simpler the spectrum. Counting the number of degrees of freedom for each of the actions studied in section \ref{seccttr} is simple: one for each scalar, one for each chiral massive vector and two for each usual massive vector. In the following, we will describe the truncated spectra in order of their increasing complexity: we will start with the S-dual dipole truncations - which describe only two propagating degrees of freedom -  then build our way up to the usual dipole (with four d.o.f), NHEK (also four) and finally the general consistent truncations, which describe six propagating degrees of freedom.

\bigskip
\noindent {\bf\emph{The S-dual dipole truncation}}

\medskip

\noindent The S-dual dipole truncation \eqref{actsddip} consists of a scalar field, $U$, and a chiral massive vector, $\hat A$, in addition to the metric. The background solution is given by \eqref{genschm} with $\l_1 =\l_2 =\l$ and $r = \rho^{-1}$. We use the linearized equations of motion \eqref{newmax}, \eqref{eqU} and \eqref{eqg} about this background to express all the components of the chiral massive vector and the metric in terms of the linearized scalar $U$, which  satisfies a fourth order equation

\bea
&&\rho^4 U^{(4)}  + 4 \rho^3 U^{(3)} - ( \k \om \rho +  2 \l^2 \k^2) \rho^2 U'' -  \k \om \rho^2 U' +\non \\ 
&& \hspace{2cm}+ \left(\l^4 \k^4 - 2 \l^2 \k^2 +  \k \om \rho (1 + \l^2 \k^2) + \frac{\k^2 \om^2 \rho^2}{4}\right) U =0 \label{equ}
\eea
The expression for the components of $\hat A$ and $g_{\mu\nu}$ in terms of $U$ is given in appendix \ref{dsdlm}. 
Asymptotically, the solution takes the form $U= \sum_i B_i \, \rho^{s_i}$, where

\be
s_{1,3} = -\half \mp \half \sqrt{1+4 \l^2 \k^2  } \;, \;\;\;\;\; s_{2,4} = \frac{3}{2} \mp  \half \sqrt{1+4 \l^2 \k^2  } \label{solsdip}
\ee
Two of these roots represent conformal dimensions and tell us about the falloff of the expectation value of the dual operator, whereas the other two are of the form $1-s$, and they tell us about the source. Thus, the holographic pairing source $\leftrightarrow$ vev should be

\be
s_1 \leftrightarrow s_4\;, \;\;\;\;\;\;\;\;s_2 \leftrightarrow s_3
\ee
One can check that these yield the expected AdS$_3$ falloffs as $\l \r 0$. Below is a plot of the four roots as a function of momentum. Interestingly, they cross over. This suggests that for $\l \k  < \frac{ \sqrt{3}}{2}$ we should consider $2s_2$ to be the dual conformal dimension, whereas for $\l \k > \frac{\sqrt{3}}{2}$, the dual conformal dimension should be $2 s_3$. It would be interesting to understand this crossover phenomenon of the $SL(2,\mathbb{R})$ weights from a holographic point of view.

\bigskip

\bigskip

\begin{figure}[h] 
\begin{center}
\includegraphics[height=4cm]{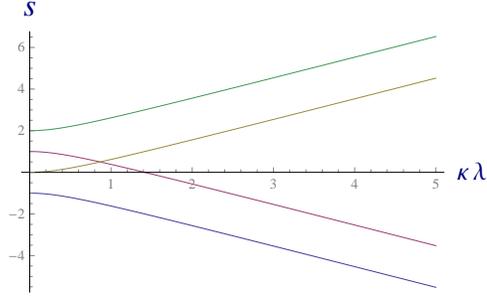} 
\caption{Plot of the four $s_k$ as a function of $\l\k$. Each solution can be identified by its value at $\l \k=0$. Note that $s_2$ and $s_3$ cross over at $\l \k = \frac{\sqrt{3}}{2}$. \label{cver}}
\end{center} 
\end{figure}

\bigskip
\noindent {\bf\emph{The usual dipole truncation}}

\medskip

\noindent The usual dipole truncation \eqref{dipact} describes four propagating degrees of freedom: two scalars, $U$ and $\phi$, and one massive vector $\hat A$. 
 Therefore, we expect to  obtain an eighth order differential equation. Nevertheless, it turns out that at least at linearized level the scalars $\phi$ and $\chi \equiv 2 U + \phi$ decouple, and each obeys a fourth order differential equation. These equations  turn out to be identical to \eqref{equ}. Thus, the spectrum of conformal dimensions of the usual dipole truncation as a function of momentum consists of two copies of that  in figure \ref{cver}.

\bigskip
\noindent {\bf\emph{The NHEK truncation}}

\medskip

\noindent In the NHEK truncation, there are four propagating degrees of freedom, so we obtain an eighth order differential equation for the function $g$, which we have chosen to be $\hat{\mathcal{A}}_v$. Plugging in $g(\rho) = \rho^s$, we find an associated weights polynomial of degree eight in $s$. This polynomial has a special form, since if we let

\be
s= \half \pm \sqrt{\b}\;, \;\;\;\;\; x \equiv \l \k
\ee
we find that $\b$ satisfies a forth order polynomial equation

\bea
&&\b^4 - (11 +12 x^2) \b^3 + (54x^4 + 79 x^2 + \frac{287}{8}) \b^2 - (108 x^6 + 177 x^4 + \frac{373}{4}x^2 + \frac{639}{16}) \b + \non \\
&& \hspace{2cm} +  \frac{2025}{256} - \frac{855 x^2}{16} + \frac{167 x^4}{8} + 117 x^6 + 81 x^8 =0 \label{eqnbeta}
\eea
We denote the four solutions  of the above equation by  $\b_i(x)$, $i=1, \ldots, 4$. As $x \r 0$, they become

\be
\b_i = \left\{ \frac{1}{4}\; , \frac{9}{4}\;, \frac{9}{4}\;, \frac{25}{4} \right\}
\ee
As $x \r \infty$, all  four solutions approach a parabola

\be\;
\lim_{x\r \infty} \; \frac{\b_i(x)}{x^2} = 3 
\ee
A plot of the four solutions is included above. Three of them - $\b_{2,3,4}$ - are monotonically increasing functions of $\l \k$, whereas the remaining one, $\b_1(x)$, first dips under zero before increasing monotonically.  The fact that $\b_1 <0$ for certain values of the momentum implies that $s$ is imaginary and the perturbation acquires non-trivial flux through the boundary of the spacetime. Such behaviour is  encountered also for  fields below the Breitenlohner-Friedman bound in AdS and may pose threats to the stability of the spacetime. Due to the oscillatory behaviour in $\rho$ near the boundary, these modes are sometimes called ``travelling waves''. We will further discuss them in the next subsection. 

\bigskip

\bigskip

\begin{figure}[h]
\begin{center}
\includegraphics[height=3.5cm]{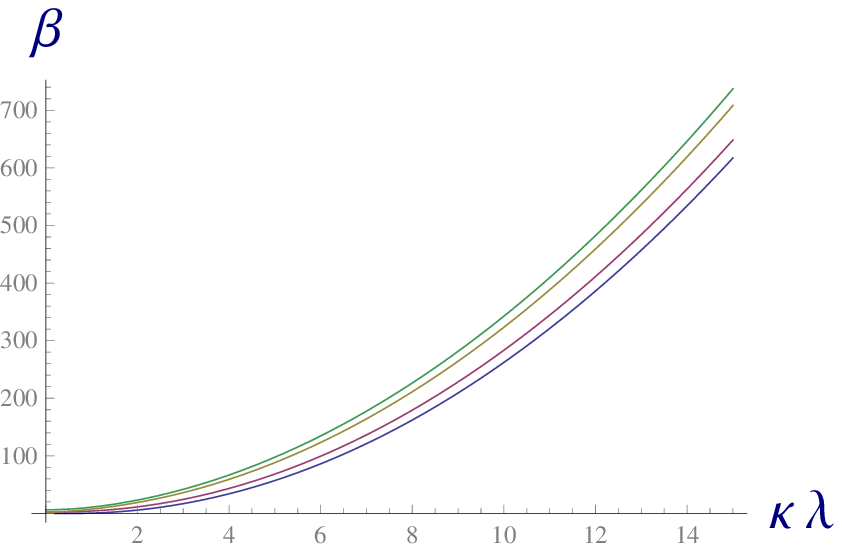} \hspace{2cm}
\includegraphics[height=3.5 cm]{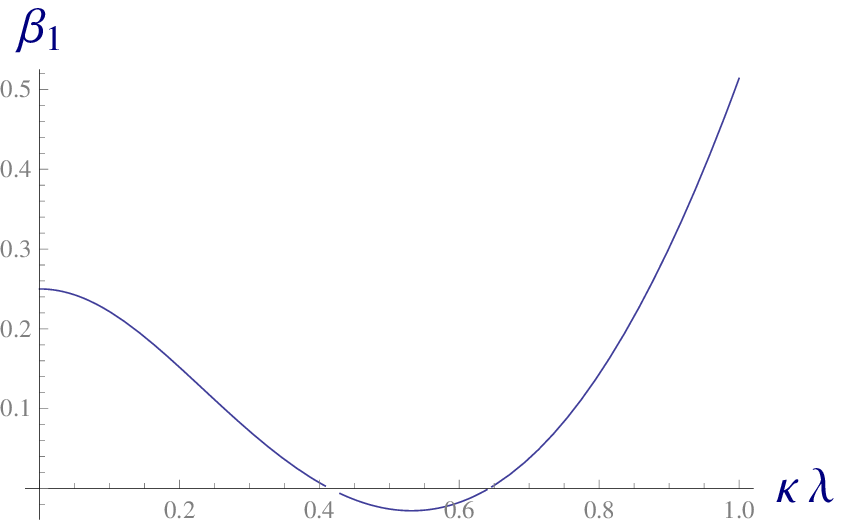}
\caption{Left: plot of the four $\b_i$ as a function of $x= \l \k$; Right: zoom near $\l \k=0$ in $\b_1$.\label{betneg}}
\end{center}
\end{figure}

\medskip

\noindent Interestingly, the eigth order differential equation we have started from can be reduced to four independent second order differential equations, all of which take the form

\be
\rho^2 g''(\rho) + \left( \frac{1}{4} - \b_i - \half \, \rho \om \k \right) g(\rho) =0 
\ee
where $\b_i$ are the four solutions to \eqref{eqnbeta}. This simplification implies that we can find the exact solution to the equations of motion in the entire spacetime, not just asymptotically, which will be of great use for the stability analysis of the next subsection.

\bigskip

\noindent {\bf\emph{The general truncation}}

\medskip

\noindent The general truncation describes six propagating degrees of freedom, and thus the weight equation is of twelveth degree. It depends on the combinations $\l_1 \k$ and $\l_2 \k$, which we can trade for two parameters  $\a$ and $x$, defined as

\be
\l_2=\a \l_1 \;, \;\;\;\;\;x= \k \l_1
\ee
The full weight equation can be again simplified by letting $s=\half \pm \sqrt{\b}$, case in which it reduces to a sixth order polynomial equation for $\b$,  written  in  appendix \ref{genwteqn}. Interestingly, this equation is symmetric under $\a \r 1-\a$, or $\l_2 \r \l_1 - \l_2 $. We can solve it with Mathematica, and plot the six solutions as a function of $\a$ and  $x$. Five of them are  positive for all $\a,x$, whereas the remaining one occasionally dips to negative values (implying an imaginary falloff) for certain values of $\a$ and $x$, plotted below

\bigskip

\begin{figure}[h] 
\begin{center}
\includegraphics[height = 5cm]{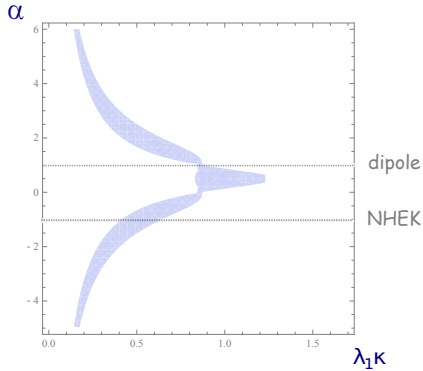}
\caption{The region in the $\a - x$ plane where the weight equation has imaginary eigenvalues. Note that the graph is symmetric about the $\a = \half$ line. The upper dotted line represents S-dual dipole theories ($\a=1$), where the range of imaginary weights has shrunk to zero size, yielding the crossover behaviour in Figure 1. The lower dotted line represents the NHEK truncation ($\a =-1$), where there is a finite range of $\k$ for which the weight is imaginary, as in Figure 2. 
}
\end{center}
\end{figure}

\medskip

\noindent Thus, we conclude that travelling waves are rather generically found in the perturbative spectrum of  these theories, where at given $\a = \l_2/ \l_1$ there exists at least some range of the momentum $\k$  for which $s$ becomes imaginary. The range of $\a$ for which these modes exist is roughly $\a \in (-5,6)$. For $\a=1$, which represents the S-dual dipole theory, the above figure shows that the range of momenta for which $s$ is imaginary  decreases to zero size. 
Thus, we can understand the crossover modes found in the dipole theory as a limiting case of the travelling wave solutions.

For $\a =0$, the Schr\"{o}dinger background is the same (up to a trivial S-duality) as that studied in \cite{Azeyanagi:2012zd} using worldsheet techniques. Some of the weights we obtain agree with a subset of the weights computed in that paper, as expected. 

\subsection{Comments on stability \label{stabanal}}

We have seen in the previous section that asymptotic falloffs involving imaginary powers of the radial distance are fairly common in the various string theory truncations. Such modes 
carry nonzero flux through the boundary at $\rho=0$, and therefore they could pose threats for the stability  of the near-horizon spacetimes. 
The question that we would like to answer in this section is: if we impose boundary conditions such that the travelling waves are confined to the interior of the spacetime, do they lead to an instability? We restrict our analysis to the classical, linearized level.  

Since the travelling wave solutions generically have non-zero flux through the boundary at $\rho=0$, energy can leak out and unitarity is lost. Nevertheless, there exists a particular linear combination of these modes for which no flux passes through the boundary; we interpret exciting this linear combination as corresponding to a physical process in which the dual theory is decoupled from exterior sources.
Together with smoothness of the solution in the interior, the zero flux  condition determines the spectrum of allowed frequencies of the normalizable modes contained within the spacetime.

This precise computation, when performed for modes below the Breitenlohner-Friedman bound in $AdS$, show that the allowed frequencies acquire a positive imaginary part, leading to an exponential blow-up  of the perturbation with time. It has been
also performed for the case of travelling waves in the $4d$ extremal Kerr near-horizon \cite{nodyn}, with the purpose of understanding the original Kerr/CFT correspondence. Again,  it was found that the zero flux condition implies a nonzero imaginary part for the frequency, which in turn leads to an instability of the four-dimensional NHEK geometry \cite{nodyn}. That $4d$ NHEK is unstable under such perturbations may seem to pose a threat for the unitarity of the dual conformal field theory. Nevertheless, one may argue that $4d$ NHEK itself corresponds to an excited state (of left-moving temperature $T_L = \frac{1}{2\pi}$) in the dual field theory, which may be allowed to  decay. 

Nevertheless, the situation is rather different here, as we find travelling wave solutions around what we believe to be the \emph{vacuum} spacetime. If these modes do trigger  an instability, then the vacuum itself is unstable. Such a result would be rather puzzling, especially in view of the fact that the $6d$ NHEK vacuum has $(0,4)$ supersymmetry\footnote{Nevertheless, it is not clear whether this amount of supersymmetry  provides a lower bound on the energy of left-movers, which is precisely the frequency $\om$ that we are concerned about.}. Luckily, there are some indications in the literature \cite{moroz}, that fields below the Schr\"{o}dinger analogue of the Breitenlohner-Friedman bound  may  in fact \emph{not} destabilize the vacuum.

In order to investigate stability we will redo the computation of \cite{nodyn} for the case of the $6d$ NHEK vacuum. We use the Schr\"{o}dinger global coordinates of \cite{hartong}

\be
\frac{1}{\ell^2} \, ds^2 = - \left( \frac{12\l^2}{\rho^2} + 1 \right)\, du^2 + \frac{2 du dv}{\rho} + \frac{d\rho^2}{4\rho^2} \label{glsch}
\ee
The Killing vectors of the global Schr\"{o}dinger spacetime \eqref{glsch} are

\be
L_{\pm 1} =  \frac{i}{2} e^{\pm 2 i  u} (\p_u + \rho \,  \p_v \pm 2 i \rho \p_\rho) \;, \;\;\;\;\; L_0 = \frac{i}{2} \p_u \;, \;\;\;\;\; K= - i \p_v
\ee
and they satisfy an $SL(2,\mathbb{R}) \times U(1)$ algebra. Note that for a perturbation with fixed frequency $\om$, the $SL(2,\mathbb{R})$ conformal weight (the eigenvalue of $L_0$) is given by $\om$. Thus, the spectrum of allowed frequencies in global coordinates gives the spectrum of allowed conformal dimensions in the holographically  dual  theory. The remaining background fields are 

\be
A = - \frac{4 \l \ell}{\rho} du \;, \;\;\;\;\; \hat A = - \frac{2\ell \l}{\rho} du
\ee
and $U=V=0$. We solve again for linearized perturbations around the global Schr\"{o}dinger background. Again, we obtain  an eighth order differential equation, which can be split as before into a series of second order ones

\be
\rho^2 f''(\rho) + \left(\frac{1}{4} -\b_i  - \frac{ \k \om \rho}{2} -\frac{ \k^2 \rho^2}{4} \right) f(\rho) =0 \label{eqnglsch}
\ee  
where now $f(\rho)$ is the scalar field $U$. The coefficients $\b_i$ are the same as before, and they satisfy the fourth order equation \eqref{eqnbeta}. We will only be interested in the solution $\b_1$ - the one that becomes negative  in figure \ref{betneg} - and only in the range of momenta in which it takes negative values

\be
\beta_1 = - \eta^2 \;, \;\;\;\;\; \eta \in \mathbb{R}^+
\ee
The solution to the  equation \eqref{eqnglsch} is the sum of a Whittaker $M$ and a Whittaker $W$ function

\be
f(\rho)=  W_{-\frac{\om}{2},i\eta} (\rho \k) + \a M_{-\frac{\om}{2},i\eta} ( \rho \k) \label{solf}
\ee
We need to pick the coefficient $\a$ in such a way that the solution is well-behaved  as $\rho \r \infty$. There are two possibilities:

\bi
\item If $\k >0$, then the combination that is smooth in the interior has $\a=0$

\item If $\k <0$, then the correct combination has

\be
\a =- \frac{i^{ \omega+1 } \Gamma\left(\frac{1-\omega}{2}+\sqrt{\beta } \right)}{\Gamma\left(1+2 i \eta\right)} \label{aimneg}
\ee
\ei
As $\rho \r 0$, the behaviour of the Whittaker functions yields

\be
f(\rho) \sim \; ( \k \rho)^{\half -i\eta} \,  \frac{ \G(2 i\eta) }{\G\left( \half + i\eta +\half \om\right)} +  (  \k \rho)^{\half +i\eta} \, \left(\a + \frac{\Gamma\left(-2 i\eta\right)}{\Gamma\left(\frac{1}{2}-i\eta+ \half \omega \right)} \right) 
\ee
One can easily check that both solutions have non-zero  symplectic flux\footnote{The flux is given by the integral of the symplectic form $\om$ over the boundary at $\rho =0$. The symplectic form $\om$ is defined as
\be
\om (\d_1 X, \d_2 X; X) = \d_2 \Theta(\d_1 X; X) - \d_1 \Theta(\d_2 X;X)
\ee
where $X$ stands for all the fields in the theory and $\Theta(\d X; X)$ is the presymplectic $d-1$ form, itself defined as the boundary term that arises in the variation of the on-shell action $\d S_{on-shell} \equiv \int d \Theta(\d X; X)$. For details see e.g.  \cite{Compere:2008us}.} through the boundary at $\rho=0$. 
After computing $\om$ for the theory \eqref{nhekact}, we demand that the flux through the surface at $\rho=0$ vanish. The condition for this is remarkably simple, and turns out to be the same as that required of a probe scalar field in $4d$ NHEK or in AdS. More precisely, if the behaviour of the solution near $\rho =0$ is

\be
f(\rho) \sim A \, \rho^{\half + i \eta} + B \, \rho^{\half-i \eta} \;, \;\;\;\;\; \sqrt{\b} \equiv  i \eta \; \;\; \mbox{for} \;\; \b <0
\ee
then the vanishing flux condition is simply 

\be
\left| \frac{A}{B} \right| =1
\ee
which is the same as the one studied in \cite{nodyn}. We distinguish two cases:

\bi
\item If $\k >0$, we have to pick the linear combination with $\a =0$ in \eqref{solf}. Then 

\be
\frac{A}{B} = \frac{(\k)^{2 i \eta} \, \Gamma\left(-2 i \eta\right) \Gamma\left(\frac{1+\om}{2}+i \eta \right)}{\Gamma\left(2 i \eta \right) \Gamma\left(\frac{1+\om}{2}-i \eta \right)} 
\ee
Note that there exists an overall ambiguity of $e^{4\pi n \eta}$ in the final answer, with $n \in \mathbb{Z}$, which is due to raising $1$ to an imaginary power. The equation we need to solve is

\be
\left| \frac{A}{B} \right| =  \frac{\left|\G\left(\frac{1}{2}+\half Re(\om)+i (\eta+Im(  \half \omega) \right)\right|}{ \left|\G\left(\frac{1}{2}+\half Re(\om)+i( \half Im( \omega) -\eta \right) \right|}=1
\ee
When $Re(\om)>0$, the function whose modulus we are taking seems monotonous as a function of
 $Im(\om)$. Consequently, the only solution to the above equation when $Re(\om)>0$ is $Im(\om)=0$.  More region plots in Mathematica show that this equation has no solution with $Im(\om) \neq 0$ if $\eta \neq 0$.  Same conclusion holds if we multiply the function by $e^{4\pi \eta n}$.

\item The second case is $\k <0$. Now, the linear combination that is smooth in the interior is \eqref{aimneg}. We can simplify\footnote{We have extensively used the identity

\be
\Gamma\left(\frac{1}{2}-i \eta+  \omega \right)\Gamma\left(\frac{1}{2}+i \eta- \omega \right) = \frac{\pi}{\cosh \pi(\eta + i \om)}\non
\ee
} the ratio of the amplitudes to

\be
\frac{A}{B} = |\k|^{2 i \eta}  \frac{ \Gamma\left(-2 i \eta\right) \Gamma\left(\frac{1-\om}{2}+i \eta \right)}{\Gamma\left(2 i \eta \right) \Gamma\left(\frac{1-\om}{2}-i \eta \right)} 
\ee
An identical analysis shows that $\om \in \mathbb{R}$ also when $\k <0$.
\ei

Since the modulus of the function

\be
f_\eta(\om) \equiv \frac{ \Gamma\left(\frac{1}{2}+i \eta+ \omega \right) }{ \Gamma\left(\frac{1}{2}-i \eta+ \omega \right)}
\ee
satisfies $|f_\eta(\om)| \in (e^{-2\pi \eta}, e^{2\pi \eta}) $,
this equation will only have solutions if we choose the branch with $n=0$. As before, the only allowed solutions have $Im(\om) =0$.

From the above analysis it seems that all real values of $\om$ are allowed by the boundary conditions. Nevertheless, requiring that the  flux through the boundary be zero \emph{at all times} $u$ also implies that, for modes of different frequencies, we must have

\be
\frac{A_1 (\om_1)}{B_1(\om_1)} \frac{A_2^\star(\om_2)}{B_2^\star(\om_2)} =1
\ee  
This implies that the allowed frequencies at fixed lightlike momentum $\k$ satisfy

\be
f_\eta(\om_1) = f_\eta(\om_2) \;, \;\;\;\;\; \om_{1,2} \in \mathbb{R}\;, \;\;\; \k >0
\ee
Given that $|f_\eta(\om)|=1$ when $\om \in \mathbb{R}$, the allowed frequencies can be obtained by intersecting the phase of $f_\eta \equiv e^{i \g_\eta(\om)}$ with a constant phase line

\bigskip
\bigskip

\begin{figure}[h]
\begin{center}
\includegraphics[height=4.5cm]{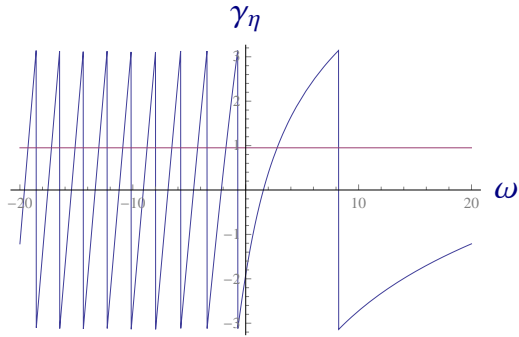} 
\caption{The spectrum of allowed $\om$ for $\k>0$ is given by the intersection of the phase function $\g_\eta(\om)$, plotted in blue above, with the constant line $\g = \g_+$, plotted in red.}
\end{center}
\end{figure}



\noindent The spectrum of allowed frequencies that we have obtained is identical to that obtained in \cite{moroz} for a probe scalar field propagating in the global Schr\"{o}dinger background. While the specific couplings in the truncated Lagrangian were extremely important in determining that travelling wave solutions exist, the behaviour of these solutions seems rather universal, and depends on the particular couplings only through the function $\eta(\k)$.

Note that the above modes do \emph{not} fall into highest weight representations of the $SL(2,\mathbb{R})$ isometry group of the background spacetime. Rather, they belong to the principal series\footnote{For this representation, the $SL(2,\mathbb{R})$ Casimir is $L^2 = j(j-1)$, with $j= \half + i s$ and $s \in \mathbb{R}$. The spectrum is given by $L_0=E_0 + n$, with $n$ an arbitrary integer and $E_0 \in (0,1)$. Note that, unlike for highest weight representations, the eigenvalues of $L_0$ are completely unrelated to the value of the Casimir. } representation \cite{Balasubramanian:1998sn}, in which there is no relationship between the Casimir of the representation and the allowed eigenvalues of $L_0$. This fact implies that there is  the imaginary falloff of the mode and the dual conformal dimension are unrelated. The spectrum is instead given by the allowed values of $\om$ that we have just computed, all of which are strictly real. Nevertheless, puzzles remain, as the spectrum of allowed $\om$ can be arbitrarily negative, and depend on one arbitrary constant, $\g_+$. It would be very interesting to understand the constraints that unitarity imposes on the conformal dimensions in the case of continuous representations. 


\section{Conclusions}

In this article we have constructed several consistent truncations of type IIB supergravity to three dimensions, which contain three-dimensional Schr\"{o}dinger solutions. These simple truncations have allowed us to compute the spectrum of perturbations about the various Schr\"{o}dinger vacua, which turned out to have rather surprising and interesting features, such as the presence of crossover modes and that of travelling waves. It would be very interesting to better understand the physics of these modes, especially since they are closely related to superradiant modes in four-dimensional NHEK.  We have also found warped analogues of the BTZ black hole that fit inside some of the consistent truncations, which should facilitate the understanding of finite-temperature holography in three-dimensional Schr\"{o}dinger spacetimes.

The presence of the travelling waves in $6d$ NHEK (and many of the other spacetimes) is rather intriguing. If the background does turn out to be stable, then these modes should have an interpretation in the holographically dual theory and one should, in principle, be able to recover the correct holographic dictionary using holographic renormalization. What is very peculiar, nevertheless, about these modes, is that their $SL(2,\mathbb{R})$ conformal dimension is unrelated to their fall-off near the boundary, and seems to depend on the boundary conditions one imposes in the deep interior of the spacetime. It would be interesting to understand how  holographic renormalization would work in this case.

Another reasonable possibility is that these modes are not stable, but this fact is not easily visible. For example, if one performs the stability analysis we have just described in AdS using \emph{null} coordinates, one does not immediately notice the Breitenlohner-Friedman instability. Since the coordinates we have used are precisely the analogue of null cordinates in AdS, one may rightfully worry that the expected instability is hidden somewhere. It would therefore be very interesting to understand it in these coordinates. One should also perform a non-linear stability analysis, as well as check for quantum instabilities of the spacetime, such as the analogue  of Schwinger pair production in $AdS_2$ \cite{Pioline:2005pf}.

Yet another possibility is that the spacetime \eqref{glsch} is not the true vacuum of the theory. Nevertheless, the analysis in \cite{moroz} shows that the unboundedness of the frequency spectrum from below is a UV effect, rather than an IR one, and one should probably expect to find it for any asymptotically Schr\"{o}dinger spacetime. Should the stability tests nevertheless hold through, one could also check whether the travelling modes lead to consistent dynamics, for example by studying whether the norm of the perturbations, given by the integral of $\om$ over a spacelike slice, is positive definite. Several such analyses have been already performed for AdS \cite{Ishibashi:2004wx,Andrade:2011dg,Andrade:2011sx,Compere:2008us} and Lifschitz spacetimes \cite{Andrade:2012xy,Keeler:2012mb}.

Let us now turn to the warped black string solutions. The fact that all solutions can be brought to a  very similar, ``warped BTZ'' form, and their thermodynamics which
resembles that of BTZ
suggest the existence of a \emph{universal} dictionary (independent of the couplings) between these black strings and finite temperature states in the dual theory. This universality is not immediately obvious; if we are to take it seriously, it suggests for example that the rather complicated combination $\tilde \l$ in \eqref{tlbh} is to be kept fixed as a function of temperature. Certainly, there is still a lot to understand in the finite-temperature Schr\"{o}dinger dictionary. Other interesting issues are: the holographic meaning of the asymptotic symmetry group analyses and the associated boundary conditions, understanding what are the correct holographic variables to use in Schr\"{o}dinger holography, the striking  resemblance \cite{scattampl} between the form of scattering amplitudes off a near-extremal black hole and thermal two-point functions in a \emph{two-dimensional} CFT. 

Another curious feature of the general solutions \eqref{genbhm} is the absence of an $SU(2) \times U(1)$ invariant consistent truncation Ansatz on the squashed $S^3$ that contains them. This fact seems to suggest that the warped black strings, which are supposed to represent finite-temperature states in a \emph{two-dimensional} theory akin to \eqref{defsrr}, live intrinsically in six dimensions. Thus, one cannot simply study them with a truncated holographic dictionary. Of course,  a simpler alternative is that we were not imaginative enough to find a consistent truncation that contains the general solutions.

The above make up only a small subset of the questions raised by the study of Schr\"{o}dinger spacetimes in string theory. We hope that the stringy consistent truncations that we have found will help yield concrete answers to at least some of them.

\bigskip

\noindent \textbf{Acknowledgements}

\medskip

\noindent We are grateful to Alejandra Castro, Geoffrey Comp\`ere, Mirjam Cvetic, Tom Hartman, James Liu, Juan Maldacena, Chris Pope, Suvrat Raju, Mukund Rangamani and Ergin Sezgin for interesting discussions. We especially thank Andrew Strominger and Balt van Rees for useful comments on the draft, and  Wei Song for collaboration in the early stages of the project.  S.D. acknowledges support from the Fundamental Laws Initiative of the Center for the Fundamental Laws of Nature, Harvard University,
and from Wallonie-Bruxelles International. The work of M.G. is supported in part by the DOE grant  DE-FG02-05ER- 41367.

\appendix

\section{Generalized spectral flows}

\subsection{The action of spectral flows \label{genform}}

In this section we review the effect of the generalized spectral flows \cite{stubena} we have discussed, namely $S\,TsT\,S$ and $T^4 S\,TsT\, ST^4$, on an arbitrary type IIB background compactified to $6d$ on $M= T^4$ and supported by purely RR three-form flux. The only requirement is that the background in question have two compact, commuting isometries, so that the metric can be written as a $T^2$ fibration over an four-dimensional base (times $T^4$)

\be
ds_{10}^2 = ds_4^2 + G_{\a\b}\, (dy^\a + \A^\a)(dy^\b + \A^\b)  + ds_M^2 \;, \;\;\;\;\; y^\a \sim y^\a + 2 \pi \label{metfib}
\ee
In addition, we assume that only the dilaton and the two-form potential are turned on, and that the $C^{(2)}$-field does not have any components in the internal directions of $M$. Thus, the latter can be decomposed as

\be
C^{(2)}_{\a\b} = \zeta \hat\e_{\a\b} \;, \;\;\;\;\; C^{(2)}_{\mu\a} = \B_{\mu\a} - C_{\a\b} \A^\b \non
\ee

\be
C_{\mu\nu}^{(2)} = \mathcal{C}_{\mu\nu} - \A_{[\mu}^\a\, \B_{\nu] \a} + \A^\a_\mu\, C_{\a\b}\,  A^\b_\nu \label{decc}
\ee
The matrix $\hat{\e}_{\a\b} = i \s_2$ represents the two-dimensional $\e$ symbol, whereas the unhatted $\e_{\a\b} = \sqrt{G}\, \hat{\e}_{\a\b}$ is the corresponding tensor density. Let us now study the effect of the first generalized spectral flow on the above generic field configuration.

\subsubsection*{The first generalized spectral flow: $S\,TsT\,S$}

After the first spectral flow, with parameter $\hat \l_1$, the metric reads

\be
ds^2_{(1)} = \Sigma_1^\half\, ds_4^2 + \Sigma_1^{-\half} G_{\a\b}\, \left(dy^\a + \A^\a_{(1)} \right)\left(dy^\b + \A^\b_{(1)}\right) + \Sigma_1^\half\, ds_M^2\label{tstmet}
\ee
The dilaton and the components of the RR two-form are givan by

\be
e^{2 \Phi^{(1)}} =  e^{2 \Phi} \Sigma_1 \;, \;\;\;\;\; \zeta^{(1)} =  \frac{\zeta + \hat \l_1 (\zeta^2 + e^{-2 \Phi} \det G)}{\Sigma_1} \non
\ee

\be
\A^\a_{(1)} = \A^\a- \hat \l_1 \, \hat\e^{\a\b} \, \B_\b\;, \;\;\;\;\;\; \B_\a^{(1)}=  \B_\a\;, \;\;\;\;\;\; \mathcal{C}_{\mu\nu}^{(1)} =  \mathcal{C}_{\mu\nu} \label{newparamone}
\ee
where we have defined

\be
\Sigma_1 =  (1+ \hat \l_1 \, \zeta)^2 + \hat \l_1^2 e^{-2\Phi} \det G
\ee
For future purposes, let us define the following  $n$-forms $H^{(n)}$ on the eight-dimensional base 

\be
H^{(1)}_{\a\b} = d\, C_{\a\b} \;, \;\;\;\;\; H^{(2)}_\a = d \B_\a - C_{\a\b}\, d \A^\b
\ee

\be
 H^{(3)} = d\, \mathcal{C} -\half\, \A^\a \wedge d \B_\a - \half \,\B_\a \wedge d \A^\a \label{defh3}
\ee
The transformation of the  forms $H^{(n)}$ under $STsTS$ is

\be
H^{(3)} \r H^{(3)} \;, \;\;\;\;\; H^{(2)}_\a \r \frac{1+ \hat \l_1 \zeta}{\Sigma_1} H^{(2)}_\a - \frac{\hat \l_1 e^{-2 \Phi} \det G}{\Sigma_1}\, \hat \e_{\a\b} d \A^\b \label{trh}
\ee
The transformation of $H^{(1)}_{\a\b} = \hat \e_{\a\b} d\zeta$ can be inferred from \eqref{newparamone}.

\bigskip

\subsubsection*{The second generalized spectral flow: $T^4S\,TsT\,ST^4$}

Now let us study the effect of the other generalized spectral flow on \eqref{metfib}-\eqref{decc}.  Let the initial background have

\be
 ds_M^2 = v(x^\mu) \sum_{i=1}^4 dx_i^2   \;, \;\;\;\;\; x^i \sim x^i + 2 \pi \label{defv}
\ee
The metric after the second spectral flow, with parameter $\hat \l_2$, is applied to a background of the type \eqref{metfib}-\eqref{decc}

\be
ds^2_{(2)} = \Sigma^{\half}_2 ds_4^2 + \S^{-\half}_2 G_{\a\b} \left(dy^\a + \A^\a_{(2)}\right)\left(dy^\b + \A^\b_{(2)} \right)  +\S^{-\half}_2 ds_M^2
\ee
where

\be
\Sigma_2 = (1+ \hat \l_2 \zeta')^2 + \hat \l_2^2 \,v^4 e^{- 2 \Phi} \det G
\ee
The Kaluza-Klein vector reads

\be
\A_{(2)}^\a = \A^\a - \hat \l_2 \hat \e^{\a\b} \B'_\b
\ee
where
\be
d \B'_\a = v^2 \, \e_\a{}^\b \star_4 d \B_\b + \zeta' \hat \e_{\a\b} d \A^\b + \frac{v^2 \zeta}{\sqrt{\det G}} \, G_{\a\b} \star_4 d \A^\b \non
\ee

\be
d \zeta' = v^2 \sqrt{\det G} \star_4 H^{(3)} \label{beach}
\ee
The above two equations determine $\zeta'$ and $\B'_\a$, up to a constant shift. The components of the  $C^{(2)}$-field and the dilaton are encoded by

\be
d \zeta^{(2)} = d\zeta \;, \;\;\;\;\; e^{2\Phi^{(2)}} = \S_2^{-1} e^{2\Phi} \label{finax}
\ee

\be
H^{(2)}_{\a\, (2)} = (1+ \hat \l_2 \zeta')\, H^{(2)}_\a + \hat \l_2 v^2 e^{-2 \Phi} \sqrt{\det G}\,  G_{\a\b} \star_4 d \A^{\b}
\ee

\be
H^{(3)}_{(2)} = \left[ (1+ \hat\l_2 \zeta')^2 - \hat \l_2^2 \, \left( e^{-2 \Phi} v^4 \det G \right) \right] H^{(3)} + \frac{\hat \l_2 (1+\hat \l_2 \zeta')}{v^2 \sqrt{\det G}} \star_4 d \left( e^{-2 \Phi} v^4 \det G \right) \label{finh3}
\ee
Alternatively, one can directly write the equation for the final $\B_\a$ -field

\be
d\B_\a^{(2)} = (1+\hat \l_2 \zeta') d \B_\a + \hat \l_2 \zeta v^2 \e_\a{}^\b \star_4 d\B_\b + \hat \l_2 v^2 \left(\zeta^2 + e^{-2\Phi} \det G \right) \, \frac{G_{\a\b}}{\sqrt{\det G}}\star_4 d\A^\b \label{finba}
\ee


\subsection{Details of generating the black hole solutions \label{horror}}

Using the above formulae, it is very easy to generate warped backgrounds in this theory by acting with STU transformations on a seed $BTZ \times S^3$ solution. 

\bigskip

\bigskip

\noindent {\bf\emph{The seed solution}}

\medskip

\noindent The seed solution \eqref{bsnull} can be viewed 
as a $T^2$ fibration with coordinates $y^\a = \{\frac{v}{R}, \half \psi\}$ over a four-dimensional base, spanned by $u,v,\th,\phi$. In the language of the previous subsection, the starting background has

\be
G_{\a\b} = \left( \begin{array}{cc} T_+^2 R^2 \ell^2 & 0 \\ 0 & \ell^2  \end{array} \right) \;, \;\;\;\;\; \A^\a = \left( \begin{array}{c} \frac{1}{ R T_+^2} r  du \\ \half \cos \th d \phi  \end{array} \right) \;, \;\;\;\;\; \B_\a = \left( \begin{array}{c} - R \ell^2 r du \\- \frac{\varepsilon \ell^2}{2} \cos \th d \phi  \end{array} \right) \label{seed}
\ee

\bigskip

\be
ds_4^2 = - \frac{ \left(r^2 - T_+^2 T_-^2\right)\ell^2 }{T_+^2} du^2 +  \frac{\ell^2 dr^2}{4 (r^2-  T_+^2 T_-^2)} + \frac{\ell^2}{4} (d\th^2 + \sin^2 \th d\phi^2)
\ee
alongside with $\Phi = \zeta= \mathcal{C}_{\mu\nu} =0$. The parameter $\varepsilon$ encodes the amount of supersymmetry preserved by the transformations, when acting on the vacuum; throughout this paper, $\varepsilon=-1$. The metric of the internal manifold is simply

\be
ds_M^2 = \sum_i  dx_i^2 \;, \;\;\;\;\; i = 1, \ldots, 4
\ee
so $v=1$ in \eqref{defv}.

\bigskip

\bigskip

\noindent {\bf\emph{The first generalized spectral flow}}

\medskip

\noindent After the first generalized spectral flow with parameter $\hat \l_1$, we have

\be
\A^\a = \left( \begin{array}{c} \frac{1}{ R T_+^2} r  du + \frac{ \hat \l_1   \varepsilon \ell^2 }{2} \cos \th d\phi \\ \half \cos \th d \phi - \hat \l_1 R \ell^2 r du \end{array} \right) \;,\;\; \;\;\;\;\; \B_\a = \left( \begin{array}{c} - R \ell^2 r du \\- \frac{\varepsilon \ell^2}{2} \cos \th d \phi  \end{array} \right) 
\ee
The dilaton, the axion and its dual become

\be
e^{2\Phi} = \S_1 \;, \;\;\;\; \zeta = \frac{\hat \l_1 R^2 T_+^2 \ell^4}{ \S_1}\;, \;\;\;\; \mathcal{C}_{\mu\nu} =0 \label{tstax}
\ee
and the metric reads\footnote{Note that the above metric is not in Kaluza-Klein form. We can bring it to such a form via the coordinate transformation $v'=v- \frac{\hat \l_1 \ell^2 R \varepsilon}{2} \, \psi$.}

\bea
ds^2 & = & \sqrt{\S_1} (ds_4^2 + ds_M^2) + \frac{T_+^2 \ell^2}{\sqrt{\S_1}} \left(dv + \frac{1}{ T_+^2}  du + \frac{\hat\l_1 \varepsilon \ell^2 R}{2} \cos \th d \phi   \right)^2 + \non \\
&& \hspace{1 cm} + \frac{\ell^2}{4 \sqrt{\S_1}} \left(d\psi + \cos \th d \phi - 2 R \ell^2 \hat \l_1 r du\right)^2
\eea
As before, we have introduced the shorthand

\be
\S_1 = 1 + \hat \l_1^2 T_+^2 R^2 \ell^4 
\ee
We find it useful to define

\be
\k = T_+^2 R^2 \ell^4  \;\;\;\; \Rightarrow \;\;\;\; \S_1 = 1 + \hat \l_1^2 \k
\ee
Note also that both the initial and final $H^{(3)} =0$.

\bigskip

\bigskip

\noindent {\bf\emph{The second spectral flow}}

\medskip

\noindent Let us now act with the second spectral flow on the geometry we obtained. First, we need to solve \eqref{beach}, whose solution reads

\be
\B'_\a =  \left( \begin{array}{c} - \varepsilon \ell^2 R r du - \frac{\hat \l_1 \k}{2} \cos \th d\phi \\
- \frac{\ell^2}{2} \cos \th d\phi + \hat \l_1 R \ell^4 r du   \end{array} \right) \;,\;\; \;\;\;\;\;\; d \zeta'=0
\ee
We choose the solution $\zeta'=0$. The new metric then reads

\bea
ds^2 & = & \sqrt{\S_1 \S_2} \, ds_4^2  +  \frac{T_+^2 R^2 \ell^2}{\sqrt{\S_1 \S_2}}  \left(d\hat v + \frac{1}{ T_+^2 R} r du + \frac{(\hat\l_1 \varepsilon + \hat \l_2) \ell^2}{2} \cos \th d \phi  - \hat \l_1 \hat  \l_2 R \ell^4  r du \right)^2 + \non \\
&+&  \frac{\ell^2}{4 \sqrt{\S_1 \S_2}} \left(d\psi + \cos \th d \phi -2 R \ell^2 ( \hat \l_1 + \varepsilon \hat \l_2) r du  - \hat \l_1 \hat \l_2 \k \cos \th d \phi\right)^2 +  \sqrt{\frac{\S_1}{\S_2}} \, dx_i^2 \non \\
&& \label{mettwospfl}
\eea
and the dilaton

\be
e^{2\Phi} = \frac{\S_1}{\S_2}
\ee
Let us now find the gauge fields. Using \eqref{finba}, it can be shown that the component $\B_\a^f$ of the final $C^{(2)}$-field satisfies the simpler equation

\be
d\B_\a^f = d\B_\a + \hat \l_2 \sqrt{\det G} \, G_{\a\b} \star_4 d\A^\b
\ee
where $\A^\a$, $\B_\a$ and $G_{\a\b}$ are those of the seed solution \eqref{seed}. We find

\be
\B_\a^f = \left( \begin{array}{c} - R \ell^2 r du  -\half \hat \l_2 \k \cos\th d\phi\\ - \frac{\varepsilon \ell^2}{2} \cos \th d \phi + \hat \l_2 \ell^4 R r du  \end{array} \right) 
\ee
The final axion satisfies \eqref{finax}, where $\zeta$ is given by \eqref{tstax}, which is a constant. Thus, we can choose $\zeta_f$ to be any constant we like. We let

\be
\zeta_f = \frac{\hat \l_1 \k}{\S_1}
\ee 
Finally, we need to compute $\mathcal{C}$. This can be done by using \eqref{finh3} to show that $H^{(3)}_f=0$, and then use the definition \eqref{defh3} to compute $\mathcal{C}$. Writing the final $C^{(2)}$ -field as

\be
C^{(2)} =\half \, C^{(2)}_{\a\b} \, dy^\a \wedge dy^b + C^{(2)}_\a \wedge dy^\a + C^{(2)}
\ee
we have

\be
C^{(2)}_\a = \frac{1}{2 \S_1}\left(\begin{array}{c} \vspace{3 mm} - 2 R \ell^2 (1 - \hat \l_1 \hat \l_2 \varepsilon \k) r du -  (\hat \l_1 + \hat \l_2) \k \cos\th d \phi \\ 2 (\hat\l_1 +\hat \l_2) R \ell^4 r  du + (\hat \l_1 \hat \l_2 \k - \varepsilon) \ell^2 \cos \th d \phi  \end{array}\right)
\ee

\be
C^{(2)}= \frac{\hat  \l_1 R \ell^4}{2 \S_1} (1-\varepsilon) (1- \hat \l_2^2 \k) \, r du \wedge\cos \th d\phi \;, \;\;\;\;\; C^{(2)}_{\a\b} = \frac{\hat \l_1 \k}{\S_1}\, \hat \e_{\a\b}
\ee

\bigskip

\bigskip

\noindent {\bf \emph{Manipulations}}

\medskip

\noindent The metric \eqref{mettwospfl} obtained after the two spectral flows is not in Kaluza-Klein form. To fix this, we  introduce the new coordinates

\be
\psi' = \frac{\psi}{1- \hat \l_1 \hat \l_2 \,\k} \;,\;\; \;\;\;\;\; v' = v -  \frac{R \ell^2(\hat \l_2 + \varepsilon \hat \l_1) \psi}{2(1-\hat \l_1 \hat \l_2\, \k)} 
\ee
Note that the original identification of the coordinate $\psi$ is \emph{not} respected by this temperature-dependent rescaling.  After this coordinate transformation, the metric takes the form

\be
ds^2  = ds_3^2 + \frac{\ell^2 \sqrt{\S_1 \S_2}}{4} d \Omega_2^2+  \frac{\ell^2 \Delta}{4 \sqrt{\S_1 \S_2}} \left(d\psi' + \cos \th\, d \phi + A\right)^2   + \sqrt{\frac{\S_1}{\S_2}} \, dx_i^2
\ee
where

\be
\Delta = (1- \hat \l_1 \hat \l_2 \,\k)^2 + \k (\hat \l_1 \varepsilon + \hat \l_2)^2
\ee

\be
A = - \frac{2 R \ell^2 (1-\varepsilon ) }{ \Delta}  (\hat \l_1 - \hat \l_2) (1- \hat \l_1 \hat \l_2 \, \k) r du +
 \frac{2 R \ell^2 T_+^2}{\Delta} (\varepsilon \hat \l_1 +\hat  \l_2) \, dv'
\ee
and the metric $ds_3^2$ in the $u,v',r$ directions will be specified later. Up to a gauge transformation, the $C^{(2)}$ field reads

\bea
C^{(2)} & = &\left[ \frac{\hat \l_1 \ell^4 R}{2 \S_1} (1-\varepsilon)  (1-\hat \l_2^2 \k)r du + \frac{\hat \l_1 + \hat \l_2}{2 R\S_1} \k dv' \right] \wedge (d\psi' + \cos \th d \phi) \non \\
&& \hspace{1cm} - \frac{\ell^2 \varepsilon}{4} (1+ \varepsilon \hat \l_2^2 \k) \cos \th d \phi \wedge d\psi' - \frac{\ell^2}{ \S_1} (1- \hat \l_1 \hat \l_2 \varepsilon \k) r du \wedge dv'
\eea
The above solution can be further embellished by performing a gauge transformation in order to make the KK gauge field descending from the metric be proportional to the one descending from 
the $C^{(2)}$ field. This can be accomplished by letting

\be
\psi' = \tilde \psi - \frac{2 (\hat \l_1 + \varepsilon \hat \l_2) R \ell^2 T_+^2}{ (1+ \varepsilon \l_1^2 \k)(1+\varepsilon \l_2^2 \k)}\, v'
\ee
Finally, we find it convenient to rescale the coordinate $v'$ and the radius $\ell$  by defining

\be
 \tilde v = \frac{(1- \hat \l_1 \hat \l_2 \k)}{(1+ \varepsilon  \hat \l_1^2 \k)(1+ \varepsilon \hat \l_2^2 \k)} \, v' \;, \;\;\;\;\;
\tilde \ell = \ell \sqrt{1+ \varepsilon \hat \l_2^2 \k}
\ee
If $\varepsilon =1$, then the final metric that we obtain is simply locally $AdS_3 \times S^3$, as expected

\be
ds^2 = \tilde \ell^2 \sqrt{\frac{\S_1}{ \S_2}} \left( T_-^2 du^2 + T_+^2 d\tilde v^2 + 2 r du d \tilde v + \frac{dr^2}{4(r^2-T_+^2 T_-^2)} + d\Om_3^2  \right)
\ee
whereas the gauge field becomes

\be
C^{(2)} = - \frac{\tilde \ell^2}{4} \cos \th d\phi \wedge d \tilde \psi -\tilde  \ell^2 \, r du \wedge d \tilde v
\ee 
This is precisely the solution we had started from, albeit with different values of the charges. For example,
   $\tilde \ell^2$, which is the flux of the Ramond-Ramond three-form field $F^{(3)}$ through the $S^3$, has changed.  This result agrees with that of \cite{stubena}, who showed that when $\varepsilon=1$, the only solution one can generate is locally $AdS_3 \times S^3$.

 What we are interested in nevertheless are the solutions with $\varepsilon =-1$, which represent nontrivial warped black strings  and are summarized below.

\bigskip

\bigskip

\noindent {\bf \emph{The final solution}}

\medskip

\noindent In the case $\varepsilon =-1$, the gauge potential takes the following simple form

\be
C^{(2)}  =  \frac{\tilde \ell^2}{4}  \cos \th d\phi \wedge d\tilde \psi -  \tilde \ell^2\, r du \wedge d\tilde v +  \frac{\tilde \ell^2  \hat \l_1 R \ell^2 }{ \S_1}  \left( r du + T_+^2 d\tilde v \right) \wedge \s_3
\ee
while the metric can be written as

\be
ds^2  =  \frac{\tilde \ell^2 \sqrt{\S_1 \S_2}}{(1-\l_2^2 \k)} \left[   \frac{(1-\l_1^2 \k)^2 (1-\l_2^2 \k)^2}{\Delta \S_1 \S_2} ds_{3}^2  + \frac{1}{4} \, d \Omega_2^2 + \frac{\Delta}{4\S_1 \S_2} \left(\s_3 +\frac{2}{\ell} A\right)^2 \right] + \sqrt{\frac{\S_1}{\S_2}} dx_i^2 
\ee
The Kaluza-Klein vector field reads

\be
A= - \frac{2 \ell (\hat \l_1 - \hat \l_2) R \ell^2 (1-\hat \l_1 \hat \l_2 \k)}{\Delta} \left( r du + T_+^2 d\tilde v \right)
\ee
and the three-dimensional part of the metric

\bea
ds_3^2  &= &   - \frac{4 \ell^4 R^2}{(1 -\hat\l_1^2 \k)^2 (1-\hat \l_2^2 \k)^2} \left[(\hat \l_1^2-\hat\l_1 \hat \l_2 + \hat \l_2^2)(1+ \hat\l_1^2\hat \l_2^2 \k^2) - \hat \l_1 \hat \l_2 \k (\hat \l_1^2 + \hat \l_2^2)\right] r^2 \, du^2 \non \\
&+&   T_+^2 d\tilde v^2 + 2 r du d\tilde  v + \frac{\Delta \S_1 \S_2}{(1-\l_1^2 \k)^2 (1-\l_2^2 \k)^2 } \left[ \frac{dr^2}{4(r^2 - T_+^2 T_-^2)} + T_-^2 du^2 \right]
\eea
The expression inside $g_{uu}$ can be simplified by defining

\be
\tilde \l =\frac{2 \ell^2 R \sqrt{\Lambda}}{(1-\hat \l_1^2 \k)(1- \hat \l_2^2 \k)}
\ee
where

\be
\Lambda = (\hat \l_1^2-\hat \l_1 \hat\l_2 + \hat\l_2^2)(1+ \hat\l_1^2 \hat \l_2^2 \k^2) -\hat \l_1 \hat\l_2 \k (\hat\l_1^2 + \hat\l_2^2)
\ee
and then noting that

\be
1+ \tilde \l^2 T_+^2 = \frac{\Delta \S_1 \S_2}{(1-\l_1^2 \k)^2 (1-\l_2^2 \k)^2 }
\ee
Consequently, the three-dimensional part of the metric takes the remarkably simple form

\be
ds_3^2 =  T_+^2 d\tilde v^2 + 2  r du d\tilde v  + \left[(1+ \tilde \l^2 T_+^2)\, T_-^2  - \tilde \l^2  r^2 \right] \,du^2 +  (1+ \tilde \l^2 T_+^2) \frac{d r^2}{4( r^2 - T_+^2 T_-^2)}  
\ee
It is perhaps remarkable that the three-dimensional part of the back hole geometries is so similar, given the rather different dependence on $\l_{1,2}$ we had in the beginning. In order to further simplify the notation, we let

\be
\l_i = \hat \l_i R \ell^2
\ee
The resulting formulae, where the tildes from $\tilde \ell$, $\tilde \psi$ and $\tilde v$ are ommitted, can be found in section \ref{nwbh}. Note that we reassign the usual identifications to the new angular coordinates, in particular $\tilde \psi \sim \tilde \psi + 4 \pi$.

\section{Details of the consistent truncations \label{cttrunc}}

\subsection{Consistent truncation for dipole theories \label{ctudip}}

As explained in the main text, we can start with the following six-dimensional action, which has been proven in \cite{duffm} to be a consistent truncation of type IIB supergravity

\be
S_{6d} =\frac{1}{16\pi G_6} \int d^6 x \sqrt{g} \left( R - \half \sum_{i=1}^2 \left[ (\p \phi_i)^2 + (\p \chi_i)^2 e^{2 \phi_i} \right] - 
\frac{1}{12} e^{-\phi_1-\phi_2} H^2 - \frac{1}{12} e^{\phi_1-\phi_2} F^2 + \chi_2 \, H \wedge F \right) \label{duffact}
\ee
where $F= dC^{(2)} + \chi_1 H$. For the dipole backgrounds we are interested in, which have $F \wedge H=0$ we can fix some of the scalars, since

\be
\chi_2 =0 \; \Rightarrow \;\; F \wedge H =0 \;, \;\;\;\;\;
\chi_1 =0 \; \Rightarrow \;\; H \wedge \star F =0 \label{vanax}
\ee 

\be
\phi_1 = \phi_2 \; \Rightarrow \;\; F^2 =0
\ee
If $F = \star F$, all three conditions are satisfied if we only require that $H \wedge F =0$.
This can be achieved by leaving $\chi_2$ in, but dropping its kinetic term. The resulting action reads

\be
S_{6d} = \frac{1}{16\pi G_6} \int d^6 x \sqrt{g} \left(R - (\p \phi)^2 - \frac{1}{12} e^{-2\phi} H^2 - \frac{1}{12} F^2 + \chi_2 H \wedge F\right)
\ee
where $\phi_1=\phi_2 \equiv \phi$. The six-dimensional equations of motion resulting from this Lagrangian are:

\be
R_{\mu\nu} = \p_\mu \phi\, \p_\nu \phi + \frac{1}{4} e^{-2\phi} H_{\mu \a\b} H_\nu{}^{\a\b} + \frac{1}{4} F_{\mu \a\b} F_\nu{}^{\a\b} 
-\frac{1}{24}\, g_{\mu\nu}  (e^{-2 \phi} H^2 + F^2 )
\ee

\be
\Box\, \phi = - \frac{1}{12} e^{-2\phi} H^2 \;, \;\;\;\;\; d (e^{-2 \phi}\star H) =0 \;, \;\;\;\;\; H \wedge F =0
\ee
and in addition $F = \star F$, $dF=dH =0$ and $\chi_2$ can be arbitrary. Note that this truncation of the IIB action resembles very
much the ten-dimensional Ansatz in \cite{maldanr}. Following \cite{malda}, we make the following Ansatz for reducing to $3d$ on a squashed three-sphere

\be
ds_6^2 = e^{-4 U - 2 V} ds_3^2 + \frac{\ell^2}{4} e^{2 U} d\Omega_2^2 + \frac{\ell^2}{4} e^{2 V} (d\psi + \cos \th d \phi)^2 
\ee

\be
B =\frac{\ell}{2} \,\hat  A \wedge (d\psi + \cos\th d\phi ) \;, \;\;\;\;\; F_3 = (1+ \star_6) f_3 \;, \;\;\;\;\; f_3 = \frac{2}{ \ell} e^{-4 (2 U + V)} \,\omega_3
\ee
where $\omega_3$ is the volume form associated to the three-dimensional metric $ds_3^2$.
The equations of motion for the scalars $U$ and $V$ read

\be
\Box U = \ell^{-2} \left(4 e^{-6 U - 2 V} - 2 e^{-8 U} - 2  e^{-8 U - 4 V} - 
 e^{-4 U - 2 \phi} \hat A^2\right) + \frac{1}{8} e^{4U - 2 \phi} \hat F^2
\ee

\be
\Box V = \ell^{-2} \left( 2 e^{-8 U} - 2  e^{-8 U - 4 V} +  e^{- 2 \phi -4 U}\hat A^2 \right)- 
\frac{1}{8} e^{-2 \phi + 4 U} \hat F^2
\ee
It is not hard to see that

\be
\Box (U+V) = 4 \ell^{-2} e^{-6U-2V} (1 - e^{-2 U -2 V})
\ee
Consequently, there exists a further consistent truncation to configurations that have $V = - U$. The three-dimensional equations of motion simplify to

\be
R_{ij} = 4 \p_i U \p_j U + \p_i \phi \p_j \phi +  \half e^{-2 \phi + 4 U} (\hat F_{i k} \hat F_j{}^k - \frac{1}{2} g_{ij} \hat F^2)   + \frac{2\, e^{-2\phi - 4 U} }{ \ell^2}\hat A_i \hat A_j + \frac{g_{ij}}{\ell^2} \left(2 e^{-8U} - 4 e^{-4U} \right)\non
\ee 

\be
\ell^2 \Box U = 2 e^{-4 U} - 2 e^{-8U} -  e^{-4U-2\phi} \hat A^2 + \frac{\ell^2}{8} e^{4U-2\phi}\hat F^2  \non 
\ee

\be
 \Box \phi = -  \frac{2}{ \ell^2} e^{-4U -2 \phi}\hat A^2 - \frac{1}{4} e^{4U-2\phi} \hat F^2 \non
\ee

\be
\nabla_i (e^{-2 \phi + 4 U}\hat F^{i k}) = \frac{4 \,e^{-2 \phi -4 U}}{\ell^2} \hat A^k \label{mxappb}
\ee
These equations can be derived from a three-dimensional action, given in \eqref{dipact}.

\subsection{Consistent truncation with RR flux \label{ct2max}}
 
Starting again from \eqref{duffact}, we make a truncation to Ramond-Ramond fields only.
 Since 

\be
\phi_1 = - \phi_2 \; \Rightarrow \;\; H^2 =0
\ee
in addition to \eqref{vanax}, we can consistently set $H=0$, 

\be
\phi_1 = - \phi_2 = \phi
\ee
and have vanishing axions.  The six-dimensional action then takes the simple form

\be
S_{6d} = \frac{1}{6\pi G_6} \int d^6 x \sqrt{g} \left( R -(\p\phi)^2 - \frac{1}{12} e^{2\phi}  F^2 \right)
\ee
with equations of motion 

\be
R_{\mu\nu} = \p_\mu \phi\,  \p_\nu \phi + \frac{1}{4}\, e^{2\phi}\, F_{\mu\a\b} F_\nu{}^{\a\b} - \frac{1}{24} \, g_{\mu\nu}\, e^{2\phi} \, F^2
\ee

\be
\Box \phi = \frac{1}{12} \, e^{2\phi} \, F^2 \;, \;\;\;\;\;\; d(e^{2\phi} \star F) =0 \label{eomphi}
\ee
The consistent truncation Ansatz  is

\be
ds_6^2 = e^{-4 U - 2 V} ds_3^2 + \frac{\ell^2}{4} e^{2 U} d\Omega_2^2 + \frac{\ell^2}{4} e^{2 V} \left(d\psi + \cos \th d \phi + \frac{2}{\ell} \, A\right)^2 \label{genctmet}
\ee

\be\,
F = \frac{2}{\ell} \, X \, \om_3+ \frac{\ell^2}{4} d\Omega_3 + \frac{\ell}{2}\, d \hat C \;, \;\;\;\;\; \hat C = \hat  A \wedge \s_3 \label{genctf}
\ee

\be
d\Omega_2 = \sin\th \, d\th \wedge d\phi \;, \;\;\;\;\; \s_3 = d\psi + \cos \th \, d\phi \;, \;\;\;\;\; d\Omega_3 = d \Omega_2 \wedge \s_3
\ee
and $\om_3$ is the volume form associated with the $3d$ Einstein metric $ds_3^2$. Note that $dF=0$ by construction. The Hodge dual of $F$ is

\bea
\star F & = & \frac{2}{\ell} \, e^{-8U-4V} \, \om_3 + e^{-4U} \,  \left(  \star_3 A -\star_3 \hat A\right) \wedge \left(\s_3 + \frac{2}{\ell} \,A \right) + \frac{\ell^2}{4} \, e^{4U} (\star_3 \hat F) \wedge d\Omega_2 \non \\
&& \hspace{1.5cm}+ \frac{\ell^2}{4}\, e^{8U+4V}  \left( X - \frac{\ell}{4} {}^{(3)} \e^{mnp} \hat F_{mn} A_p\right)  \, \left(d\Omega_3 +\frac{2}{\ell} \, d\Omega_2 \wedge A\right)
\eea
The Maxwell equation $d(e^{2\phi} \star F) =0$ implies that the coefficient of $d\Om_3$ in $e^{2\phi} \star F $ must be constant. Thus,

\be
X = e^{-8U-4V-2\phi} + \frac{\ell}{4} {}^{(3)} \e^{mnp} \hat F_{mn} A_p
\ee
where we have chosen the unfixed constant by the requirement that $AdS_3$ of radius $\ell$ be a solution. It also gives the following $3d$ Maxwell equation

\be
d(e^{4U + 2 \phi} \star_3 \hat F) + \frac{2}{\ell} \, F + \frac{4}{\ell^2} \, e^{-4U + 2 \phi} \left(\star_3 \hat A -  \star_3 A \right) =0 \label{eomfhat}
\ee
Let us also compute

\be
F_6^2 = \frac{24}{\ell^2} \, e^{-4U -2V} (1-e^{-4\phi}) + \frac{24}{\ell^2} \, e^{2V} (\hat A_i - A_i)(\hat A^i -  A^i) + 3\, e^{8U+2V} \hat F_{ij} \hat F^{ij}
\ee
The dilaton equation reads

\be
\Box_3 \phi= \frac{2}{\ell^2} \, e^{-8U-4V+2\phi} (1-e^{-4\phi}) + \frac{2}{\ell^2}\, e^{-4U + 2\phi} \, (\hat A -  A)^2 + \frac{1}{4} \, e^{4U + 2\phi} \, \hat F^2 \label{eomphi}
\ee 
Let us now move on to Einstein's equations. We have 

\bea
R_{i\psi} - A_i R_{\psi\psi} &=& - \frac{\ell}{4} \, e^{4V+4U} (\nabla_k F^k{}_i + 4 F^k{}_i \p_k(U+V))\non \\
& = & \frac{1}{4} \, \e_{ijk} \hat F^{jk} +  \frac{1}{\ell} \, e^{2\phi-4U} \left(\hat A_i - A_i\right)
\eea
which implies the following Maxwell equation for the Kaluza-Klein vector field

\be
\nabla_k (e^{4U+4V} F^k{}_i) + \frac{1}{\ell} \, \e_{ijk}\, \hat F^{jk} + \frac{4}{\ell^2} \,e^{2\phi-4U} \left( \hat A_i - A_i\right) =0 
\ee
The two Maxwell equations, combined, yield

\be
 e^{4U + 2 \phi} \star_3 \hat F + \frac{2}{\ell} \, \hat A =- e^{4U+4V} \star_3 F - \frac{2}{\ell} A \label{constrtwoas}
\ee
up to a $d$-exact term. The equations of motion for $U$ and $V$ are

\bea
\Box V & = & \frac{1}{\ell^2} \, e^{-8U} (2-e^{-4V+2\phi} -e^{-4V -2\phi}) + \frac{1}{4} \, e^{4U+4V}  F^2 + \frac{1}{\ell^2}\, e^{-4U+2\phi} \left( \hat A - A\right)^2 - \frac{1}{8} \, e^{4U+2\phi} \, \hat F^2 \non \\
\Box U & = & \frac{1}{\ell^2} \, e^{-8U} (4 e^{2U-2V} - 2 - e^{2\phi-4V} -e^{-2\phi-4V}) - \frac{1}{\ell^2}\, e^{-4U+2\phi} \left( \hat A - A\right)^2 + \frac{1}{8} \, e^{4U+2\phi} \, \hat F^2 \non \\ &&
\eea
The $3d$ Einstein equation reads

\bea
{}^{(3)}R_{ij} &=& 6 \p_i U \p_j U + 2 \p_i V \p_j V + 2 (\p_i U \p_j V + \p_j U \p_i V)+ \p_i \phi \, \p_j\phi + \frac{1}{2} \, e^{4U+4V} \left(F_{ik}F_j{}^k - \half F^2 g_{ij} \right) + \non \\
&+&  \frac{1}{2} \, e^{4U+2\phi} \left(\hat F_{ik} \hat F_j{}^k - \half \hat F^2 g_{ij} \right) + \frac{2}{\ell^2} \, e^{-4U+2\phi} (\hat A_i -A_i)(\hat A_j -A_j) - \non \\
&-& \frac{2}{\ell^2} \, g_{ij} \,e^{-8U} \left(4\, e^{2U-2V} -1 - e^{2\phi-4V} -  e^{-2\phi-4V} \right)
\eea
The effective $3d$ action is then

\bea
S_{3d} &=& \int d^3x \sqrt{g} \left[{}^{(3)}R - 6 (\p U)^2 -2(\p V)^2 - 4 \p_i U \p^i V - (\p \phi)^2 - \frac{1}{4} \,e^{4U+4V} \,F^2 - \frac{1}{4} \, e^{4U+2\phi} \hat F^2  \right. \non \\
&& \left. - \frac{2}{\ell^2} \, e^{-4U+2\phi} (\hat A-A)^2 + \frac{2}{\ell^2} \, e^{-8U}  \left(4\, e^{2U-2V} -1 - e^{2\phi-4V} -  e^{-2\phi-4V} \right) + \frac{1}{\ell}\, \e^{ijk} A_i \hat F_{jk}\right] \non \\ &&
\eea
and it is not hard to check that it yields all the equations of motion we have written so far.

\subsection{Consistent truncation for S-dual dipoles \label{cttrsddip}}

The Ansatz for the $3d$ fields is \eqref{genansatz} with, additionally, $A=0$. Einstein's equations yield the following equations for  $\hat A$, which follows from the fact that  $R_{ia} =0$

\be
F_{i\,\a\b} F_a{}^{\a\b} =0  \;\;\;\;\; \Rightarrow \;\;\;\;\; \hat F_{ij} = \frac{2}{\ell} e^{2
\phi -4U} \e_{ijk} \hat A^k \label{newmax}
\ee
This implies that \eqref{eomphi}, written in terms of $3d$ variables, reads 

\be
\Box \phi = \frac{2}{\ell^2} \, e^{-8U-4V + 2 \phi} (1-e^{-4\phi}) +  \frac{2}{\ell^2} \, e^{-4U+2\phi} \hat A^2 (1-e^{4\phi})
\ee
which means we can safely truncate to backgrounds where $\phi=0$. The above expression is proportional to $F^2$, which we from now on will set to zero inside Einstein's equation. 
The above equation \eqref{newmax} for the vector field is perfectly consistent with the Maxwell equation derived from $d(e^{2\phi} \star F) =0$, provided that $\phi=0$. The $\psi\psi$ and $\th\th$ components of the Einstein equations can be  combined to yield

\be
\frac{\ell^2}{4} \,\Box (U+V) = e^{-6U-2V} (1-e^{-2U-2V})
\ee
which shows that we can further truncate the Ansatz to $V=-U$.  After this truncation, the equation for $U$ reads

\be
\frac{\ell^2}{2} \, e^{4U} \Box U = 1 - e^{-4U} - \hat A^2 \label{eqU}
\ee
Finally, the $3d$ Einstein equation reads

\be
{}^{(3)} R_{ij} + \frac{2}{\ell^2} \, e^{-4U}\, (2-e^{-4U}) \, g_{ij} = 4 \p_i U \p_j U  +\frac{4}{\ell^2} \, e^{-4U} \hat A_i \hat A_j \label{eqg}
\ee
These equations of motion can be again derived from a three-dimensional action, which is given in \eqref{actsddip}.

\subsection{Consistent truncation for NHEK}

We start with the same consistent truncation Ansatz as in the previous subsection, \eqref{genctmet}-\eqref{genctf}.
The  NHEK background has $F=\star F$ and $\phi=0$, conditions which are compatible with each other, since 

\be
\phi_1 = \phi_2\, (=0) \;\;\;\;\; \Rightarrow \;\;\;\; F^2 =0
\ee
Imposing the self-duality condition $F=\star F$ we obtain

\be
\frac{2}{\ell} ( \hat A -  A) =  e^{4U} \star_3 \hat F \;, \;\;\;\;\;\; 
X = e^{-8U-4V} + \frac{\ell}{4} {}^{(3)} \e^{mnp} \hat F_{mn} A_p
\ee 
Next, Einstein's equations impliy the following  equation for the Kaluza-Klein vector field

\be
 d(e^{4U+4V} \star_3 F) + \frac{2}{\ell}\, \hat F - \frac{4}{\ell^2}  \,e^{-4U} \star_3  \left( \hat A - A\right)=0 \label{maxkk}
\ee
Combining the two Maxwell  equations, we find

\be
e^{4U+4V} \star_3 F= - \frac{4}{\ell} \hat A
\ee
Using the above relations,  the equations of motion for $U$ and $V$ simplify to

\be
\Box U = \frac{2}{\ell^2} \, e^{-8U} (2 e^{2U-2V} - 1 - e^{-4V}) - \frac{2}{\ell^2} \, e^{-4U} \left(\hat A - A\right)^2
\ee

\be
\Box V = \frac{2}{\ell^2} \, e^{-8U} (1- e^{-4V}) -\frac{8}{\ell^2}\, e^{-4U -4V} \hat A^2 + \frac{2}{\ell^2} \, e^{-4U} \left(\hat A - A\right)^2
\ee
whereas the Einstein equation becomes

\bea
{}^{(3)}R_{ij}&=& 6 \p_i U \p_j U + 2 \p_i V \p_j V + 2 (\p_i U \p_j V + \p_j U \p_i V)+ \frac{8}{\ell^2} \,e^{-4U-4V} \, \hat A_i \hat A_j + \non \\
& + & \frac{4}{\ell^2}\, e^{-4U} \, (\hat A_i -A_i)(\hat A_j-A_j) - \frac{2}{\ell^2} \, g_{ij} \,e^{-8U} \left(4\, e^{2U-2V} -1 -2 e^{-4V}  \right) 
\eea
The above equations can be derived from a three-dimensional action, given in \eqref{nhekact}. 

\section{The Kaluza-Klein reduction}

In this appendix we compute the components of the Christoffel symbols and the Ricci tensor associated with the truncation\footnote{For simplicity, we have rescaled by $\ell/2$ the KK vector field.}

\be
ds^2_6 = e^{-4 U - 2 V}  ds^2_3 + \frac{\ell^2}{4} \,e^{2 U} (d\th^2 + \sin^2 \th d\phi^2) + \frac{\ell^2}{4} \, e^{2 V} (d \psi + \cos \th d \phi + A)^2
\ee
At intermediate steps, it will be useful to use the conformally-transformed three-dimensional metric

\be
d\tilde{s}^2 = \Om^2 ds_3^2 \;, \;\;\;\;\; \Om^2 = e^{-4U -2V}
\ee
The inverse metric in the block-diagonal $\{x^\mu, \phi, \psi \}$ part reads

\be
g^{MN} = \left(\begin{array}{ccc} \vspace{5 mm}\;\;\;\; \tilde{g}^{\mu\nu} \;\;\;\; & 0 & - \tilde{A}^\nu\\\vspace{0.5 cm } 0 &  \frac{4 e^{-2 U} }{\ell^2 \sin^2 \th} & -   \frac{4 e^{-2 U} \cos \th}{\ell^2\sin^2 \th} \\  - \tilde{A}^\mu &  -   \frac{4 e^{-2 U} \cos \th}{\ell^2\sin^2 \th} \;\;\; &\;\; \tilde{A}^2 + \frac{4}{\ell^2} (e^{-2 V}  + e^{-2U} \cot^2 \th )   \end{array} \right)
\ee
The Christoffel symbols are now

\be
{}^{(6)} \Gamma^\l_{\mu\nu} = {}^{(3)} \tilde \Gamma^\l_{\mu\nu} + \half\,\frac{\ell^2}{4}\, \tilde g^{\l \rho} e^{2 V} (A_\nu F_{\mu\rho} + A_\mu F_{\nu \rho} - 2 \p_\rho V A_\mu A_\nu) 
\ee

\be
\Gamma^\psi_{\mu\nu} = \half (\tilde \nabla_\mu A_\nu + \tilde \nabla_\nu A_\mu) - \half\,\frac{\ell^2}{4}\, e^{2 V} \tilde{A}^\l (A_\nu F_{\mu\l} + A_\mu F_{\nu\l}) +\frac{\ell^2}{4}\, e^{2V} A_\mu A_\nu \tilde A^\l \p_\l V + A_\mu \p_\nu V + A_\nu \p_\mu V 
\ee

\be
\Gamma^\l_{\psi \mu} =\frac{\ell^2}{4} e^{2V} \left( \half\, \tilde g^{\l\rho} F_{\mu\rho}- A_\mu \tilde{g}^{\l\rho} \p_\rho V \right)\;, \;\;\;\;\;\; \Gamma^\l_{\phi \mu} = \Gamma^\l_{\psi \mu}  \cos \th
\ee
Note that the Kaluza-Klein vector field does not contribute to symbols of the form $\Gamma^\l_{ab}$

\be
\G^\l_{\th\th} = - \frac{\ell^2}{4} \,\Om^{-2} e^{2U} \p^\l U \;, \;\;\;\;\; \G^\l_{\psi\psi} = - \frac{\ell^2}{4}\,\Om^{-2} e^{2V} \, \p^\l V \;, \;\;\;\;\; \G^\l_{\psi\phi} = -\frac{\ell^2}{4}\,\Om^{-2} e^{2V} \cos\th\,  \p^\l V
\ee

\be
\G^\l_{\phi\phi} = -\frac{\ell^2}{4}\, \Om^{-2} (\sin^2 \th e^{2U} \p^\l U + \cos^2 \th e^{2V} \p^\l V)
\ee
Next,

\be
\Gamma^\th_{\l \th} = \Gamma^\phi_{\l \phi} = \p_\l U \;,\;\;\;\;\;\Gamma^\th_{\l \phi} = \half \, e^{2V-2U} A_\l \sin\th \;, \;\;\;\;\; \Gamma^\phi_{\l \th} = - \frac{1}{2\sin\th} \, e^{2V-2U} A_\l 
\ee

\be
\Gamma^\psi_{\l \th} = \half \,e^{2 V-2 U} A_\l \cot \th \;, \;\;\;\;\; \Gamma^\psi_{\l \psi} = \p_\l V - \half \,\frac{\ell^2}{4}\, \tilde A^\rho F_{\l\rho} e^{2V} + \frac{\ell^2}{4}\,e^{2V}  A_\l \tilde A^\rho \p_\rho V
\ee

\be
\Gamma^\psi_{\l \phi} = (\p_\l V-\p_\l U) \cos \th -\half\,\frac{\ell^2}{4} \, e^{2V} \tilde A^\rho F_{\l \rho} \cos \th +\frac{\ell^2}{4}\, A_\l \tilde A^\rho \p_\rho V e^{2V} \cos\th
\ee
The $\Gamma^a_{bc}$ are the same as the ones without a gauge field,

\be
\G^\th_{\phi\phi} = (e^{2V-2U} -1) \sin \th \cos \th \;, \;\;\;\; \G^\th_{\phi\psi} = \half \, e^{2V-2U} \sin \th \;, \;\;\;\;\G^\phi_{\th \phi} =  (1 - \half \,e^{2V-2U} ) \cot \th
\ee

\be
 \G^\phi_{\th \psi} = -\frac{e^{2V-2U}}{2 \sin \th } \;, \;\;\;\; \G^\psi_{\th \psi} = \frac{ e^{2V-2U}}{2} \cot \th \;, \;\;\;\; \G^\psi_{\th \phi} = - \half \sin \th + \frac{\cos^2\th}{\sin \th} \left(\frac{ e^{2V-2U}}{2} -1\right) 
\ee
plus the additional ones

\be
\Gamma^\psi_{\th\th} =\frac{\ell^2}{4}\, \tilde A^\l \p_\l U e^{2U} \;, \;\;\;\;\; \Gamma^\psi_{\phi\phi}=\frac{\ell^2}{4}\, \tilde A^\l (\p_\l U e^{2U} \sin^2\th + \p_\l V e^{2V} \cos^2\th) 
\ee

\be
\Gamma^\psi_{\phi\psi} = \frac{\ell^2}{4}\, \tilde A^\l \p_\l V e^{2V} \cos \th \;, \;\;\;\;\; \Gamma^\psi_{\psi\psi} = \frac{\ell^2}{4}\,\tilde A^\l \p_\l V e^{2V} 
\ee
Perhaps surprisingly, the contracted Christoffel symbols $\Gamma^M_{M N}$ do not depend on the vector field

\be
\Gamma^M_{M \l} = {}^{(3)} \tilde \Gamma^\rho_{\rho\l} + 2 \p_\l U + \p_\l V \;, \;\;\;\;\; \Gamma^M_{M\th} =\cot \th \;, \;\;\;\;\;\; \Gamma^M_{M \phi} =\Gamma^M_{M\psi} =0 
\ee
Using these formulae one can show that 

\be
\Box_6 \phi  = e^{4U +2V} \Box_3 \phi
\ee

\bigskip

\noindent We now compute the components of the Ricci tensor. We find

\be
R_{i\th} = R_{\th\phi} = R_{\th\psi} =0 
\ee

\be
R_{\th\th} = 1-\half \,e^{2V-2U} - \frac{\ell^2}{4} e^{2U} \Om^{-2}  \Box U 
\ee

\be
R_{\psi\psi} = - \frac{\ell^2}{4} \Omega^{-2} e^{2V} \Box V + \frac{1}{2} e^{4V-4U} + \frac{\ell^4}{64} \, \Omega^{-4} \, e^{4V} \, F^2
\ee
\bea
R_{\psi \, i} &= &\half A_i e^{4V-4U} - \frac{\ell^2}{4} e^{2V} \Om^{-2} A_i \Box V + \frac{\ell^4}{64} e^{4V} \Om^{-4} A_i F^2 - \frac{\ell^2}{8} e^{2V} \Om^{-2} \nabla_k F^k{}_i\non \\
&& \hspace{7cm} + \frac{\ell^2}{2} e^{2V} \Om^{-2} F_i{}^k \p_k( U +  V)
\eea
\bea
R_{\phi\phi}  &=&  -  \Omega^{-2} \frac{\ell^2}{4} \,\left( e^{2 U} \Box U \sin^2 \th + e^{2 V}  \Box V  \cos^2 \th\right) +
\sin^2 \th + \half e^{4 (V - U)} \cos^2 \th - \half e^{2 (V - U)} \sin^2 \th \non \\
 && \hspace{3cm} + \frac{\ell^4}{64} \, e^{4V} \, \tilde F^2\, \cos^2\th 
\eea

\bea
R_{ij} &=& R_{ij}^{(0)} + \half A_i A_j e^{4V-4U} - \frac{\ell^2}{4} \Om^{-2} e^{2V} A_i A_j  \Box V + \frac{\ell^4}{64} e^{4V} \Om^{-4}  F^2 A_i A_j +  \frac{\ell^2}{8} e^{2V} \Om^{-2} F_{ik}  F^k{}_j - \non \\
&-&  \frac{\ell^2}{8} e^{2V} \Om^{-2} (A_i \nabla_k  F^k{}_j + A_j  \nabla_k  F^k{}_i) +  \frac{\ell^2}{2} e^{2V} \Om^{-2} [A_i  F_j{}^k \p_k (U+V) + A_j F_i{}^k \p_k (U+V)] \non \\&&
\eea
where $R_{ij}^{(0)}$ is the part of the Ricci tensor that does not depend on the Kaluza-Klein vector field. It is given by
\be
R_{ij}^{(0)} = {}^{(3)}R_{ij} - 6 \p_i U \p_j U - 2\p_i V \p_j V - 2 (\p_i U \p_j V + \p_i V \p_j U) + 
g_{ij} (2 \Box U + \Box V) 
\ee
Finally,

\be
R_{\phi\psi} = R_{\psi\psi} \cos \th \;, \;\;\;\;\; R_{\phi \, i} = R_{\psi \, i} \, \cos\th
\ee

\section{Details of the linearized modes}

\subsection{Dipole S-dual solution \label{dsdlm}}

We expand the fields around the background S-dual dipole Schr\"{o}dinger solution

\be
ds^2 =\ell^2 \left( - \frac{4\l^2\, du^2}{\rho^2} + \frac{2 du dv}{\rho} + \frac{ d\rho^2}{4\rho^2}\right)\;, \;\;\;\;\; \hat A=- \frac{2 \l \ell \, du}{\rho}
\ee
and solve the linearized equations of motion \eqref{newmax}, \eqref{eqU} and \eqref{eqg} with Mathematica. The Fourier space solution for the $X$ modes reads

\bea
\hat{ \mathcal{A}}_u^X & = & -\frac{2\ell}{ \k^2 \l \rho } \left( 2 U+\l^2 \kappa^2   U+\frac{1}{2}  \rho \kappa  \omega  U+\frac{\l^2 \k^2}{4}  \rho \kappa \omega  U+\frac{1}{8} (\rho \kappa\omega)^2 U-2 \rho\, U'-\l^2 \k^2  \rho \, U'- \right. \non \\
&& \hspace{3cm} \left.  - \frac{1}{2}  \rho^2 \kappa  \omega \, U'+\rho^2 U''-\frac{1}{4}  \rho^3 \kappa  \omega \, U''+\rho^3 U^{(3)} \right)
\eea

\be
\hat{\mathcal{A}}_v^X =-\frac{\ell}{2\l} \left( 2 U +\l^2 \kappa ^2   U+\frac{ \rho \kappa  \omega  U}{2 }-\rho^2 U'' \right)
\ee

\be
\hat{\mathcal{A}}^X_\rho =  \frac{i \ell}{2 \k \l \rho} \left(\rho^3 U''' + \rho^2 U'' - \half  \k \om \rho^2 U' - \l^2 \k^2 \rho U' - 2 \rho U' + \l^2 \k^2 U + 2 U  \right)
\ee

\be
h_{uu}^X = \frac{8 \ell^2}{ \k^2 \rho^2} \left( \rho^2 U'' - 2 U - \l^2 \k^2 U - \half \, \k \om \rho U \right)
\ee
The remaining components of $h_{ij}$ do not receive an $X$-mode contribution.

\subsection{The general weight equation \label{genwteqn}}

We similarly expand all fields of the consistent truncation \eqref{gentract} at linearized level around the background solution given by \eqref{genbckm}, \eqref{genbcka} and vanishing scalars. Using Mathematica, we  express all fields in terms of a particular combination of $U$ and $\phi$, given by

\be
\xi \equiv \frac{\l_1+\l_2}{2\l_1}\, U - \frac{\l_1-\l_2}{4\l_1} \,\phi
\ee
As expected, we find that $\xi$ satisfies a twelveth order differential equation which, interestingly, is symmetric under $\a \r 1-\a$, where $\a = \l_2/\l_1$. Next, plugging in $\xi = \rho^s$ and collecting the leading powers as $\rho \r 0$, we find a polynomial equation for $s$. This equation can be further symplified by letting

\be
s= \half \pm \sqrt{\b}
\ee
and noting that for either sign, the twelveth order equation reduced to a sixth order equation for $\b$. This equation  reads

\bea
&&\hspace{-1cm} \beta ^6-\frac{3}{2} \left(9+x^2 \tilde \alpha^2\right) \beta ^5+\frac{1}{16} \left(1023+238 x^2 \tilde\alpha^2+15 x^4 \tilde\alpha^4\right) \beta ^4+\frac{1}{16} \left(-2173-735 x^2\tilde\alpha^2-103 x^4\tilde\alpha^4-5 x^6 \tilde\alpha^6\right) \beta ^3+\non\\
&& \hspace{-1cm} + \frac{1}{256} \left(32751+12572 x^2 \tilde\alpha^2+2938 x^4 \tilde\alpha^4+348 x^6 \tilde\alpha^6+15 x^8 \tilde\alpha^8\right) \beta ^2+\frac{1}{512} \left(-21627-4527 x^2 \tilde\alpha^2-1838 x^4 \tilde\alpha^4 \right. \non \\
&& \hspace{-1cm} \left.-574 x^6 \tilde\alpha^6-71 x^8 \tilde\alpha^8-3 x^{10} \tilde\alpha^{10}\right) \beta +\frac{1}{4096}\left(18225-810 x^2 \tilde\alpha^2-2961 x^4 \tilde\alpha^4+127 x^8 \tilde\alpha^8+22 x^{10} \tilde\alpha^{10}+x^{12} \tilde\alpha^{12}\right.\non \\
&&\hspace{-1cm} \left. -4 x^6 \left(65536-32768 \tilde\alpha^2+4096 \tilde\alpha^4+51 \tilde\alpha^6\right)\right)=0
\eea
We have used the shorthand

\be
\tilde \a = 2 \sqrt{1-\a + \a^2}
\ee
Note that $x\tilde \a = \tilde \l \k$ and that most of the wave equation depends on only this combination, except for the terms on the last line.

\end{document}